\documentclass[prc,nofootinbib,amsmath,amssymb,superscriptaddress,groupedaddress,showpacs]{revtex4-1}
\usepackage{epsfig}
\usepackage{dcolumn}
\usepackage{bm}
\usepackage{amssymb}
\usepackage{amsmath}
\usepackage{multirow}
\usepackage [english]{babel}
\usepackage [autostyle, english = american]{csquotes}
\MakeOuterQuote{"}
\usepackage{array}
\usepackage[colorlinks=true]{hyperref}
\usepackage[usenames,dvipsnames,table]{xcolor}
\usepackage{color}
\usepackage[shortlabels]{enumitem}
\usepackage{rotating}

\begin{document}

\title{Multi-particle correlations, cumulants, and moments sensitive to fluctuations in rare-probe azimuthal anisotropy in heavy ion collisions}

\author{Abraham Holtermann\footnote{Email: ath8@illinois.edu}}
\author{Jacquelyn Noronha-Hostler \footnote{Email: jnorhos@illinois.edu}}
\author{Anne M. Sickles \footnote{Email: sickles@illinois.edu}}
\author{Xiaoning Wang \footnote{Email: xw31@illinois.edu}}

\affiliation{\small{\it Department of Physics, University of 
Illinois, Urbana, IL 61801, USA}}

\date{\today}
\begin{abstract}

Correlations of two or more particles have been an essential tool
for understanding the hydrodynamic behavior of the quark-gluon plasma created
in ultra-relativistic nuclear collisions.
In this paper, we  extend that framework to introduce a mathematical construction of multi-particle correlators that utilize correlations between arbitrary numbers of particles of interest (e.g. particles selected for their strangeness, heavy flavor, and conserved charges) and 
inclusive reference particles to estimate the azimuthal anisotropies of rare probes. To estimate the fluctuations and correlations in the azimuthal anisotropies of these particle of interest, we use these correlators in a system of cumulants, raw  moments, and central moments. We finally introduce two classes of observables that can compare the fluctuations in the azimuthal anisotropies of particles of interest with reference particles at each order.
  
\end{abstract}
\newcommand{\qhat}{\mbox{$\hat{q}$}}
\newcommand{\vn}{\mbox{$v_{n}$}}
\newcommand{\vnprime}{\mbox{$v'_{n}$}}
\newcommand{\Psin}{\mbox{$\Psi_{n}$}}
\newcommand{\nth}{\mbox{$n$-th}}
\newcommand{\vtwo}{\mbox{$v_{2}$}}
\newcommand{\pt}{\mbox{$p_{T}$}}
\newcommand{\vthree}{\mbox{$v_{3}$}}
\newcommand{\vfour}{\mbox{$v_{4}$}}
\newcommand{\avg}[1]{\left\langle #1 \right\rangle}
\renewcommand{\thefootnote}{\fnsymbol{footnote}}
\maketitle

\section{Introduction}

 The quark-gluon plasma (QGP)  is a state of hot nuclear matter
 created in the collisions of nuclei at the Large Hadron Collider (LHC)
 and the Relativistic Heavy Ion Collider (RHIC); for a recent 
 review see Ref.~\cite{Busza:2018rrf}.
 The  fluid nature of this matter has been confirmed using collective flow measurements at both colliders~\cite{Heinz:2013th,Luzum:2013yya,DerradideSouza:2015kpt}. The early stages of heavy ion collisions produce predominately elliptical shapes due to the nature of the geometry of these collisions but other geometrical shapes are possible due to quantum mechanical fluctuations of the quarks and gluons within the nucleus.  Due to the asymmetric pressure gradients caused by these geometrical shapes, the fluid nature of the QGP converts these geometrical shapes into collective flow patterns in momentum space. These collective properties manifest as azimuthal anisotropies in the distribution of particle angles $\phi$, and are quantified by $v_n$, the Fourier harmonics for the distribution of particle yields around $\Psi_n$, the n-th order event-plane:   
\begin{equation}\label{eq:flowgeneral}
\frac{d N}{d \phi} \propto 1+2 \sum_{n=1}^{\infty} v_n \cos \left[n\left(\phi-\Psi_n\right)\right].
\end{equation}
where non-zero harmonics $v_n$ attest to collectivity within the strongly interacting, nearly perfect fluid nature of the QGP.
The cumulant framework~\cite{Borghini:2000sa,Borghini:2001vi,Bilandzic:2010jr,Bilandzic:2013kga} has also been used extensively to measure \vn~\cite{ATLAS:2014qxy,ALICE:2014dwt,CMS:2013jlh,STAR:2015mki,ATLAS:2017rtr,CMS:2017glf,ALICE:2019zfl,ATLAS:2019peb}.
One particular advantage of using cumulants is that these measurements can provide sensitivity 
to both the root mean square (RMS) values of the anisotropies, $\sqrt{\langle \vn^2 \rangle}$, over some collection of events, but
also to higher order fluctuations present in the distribution of event-by-event \vn.
Measurements of \vn\ are essential for constraining transport coefficients within relativistic viscous fluid dynamics using Bayesian analyses \cite{Bernhard:2019bmu, JETSCAPE:2020mzn,Nijs:2020ors,Parkkila:2021tqq,Heffernan:2023utr} and are one of the  standard benchmarks that  models must reproduce~\cite{Alba:2017hhe,Schenke:2020mbo}. 
Previous work has demonstrated that experimental measurements of flow fluctuations  \cite{ALICE:2011ab,CMS:2012zex,ATLAS:2017hap,ATLAS:2019peb,STAR:2022gki} can play a crucial role in constraining initial state models \cite{Renk:2014jja,Giacalone:2017uqx,Carzon:2020xwp}.

Up until this point we have discussed generic collective flow harmonics that are measured using  a nearly inclusive set of particles dominated by those at low transverse momenta ($p_T$), \textit{reference particles}.  However, a number of useful signals of the QGP come from Particles of Interest (POI), a class of either identified particles (e.g. protons) or high $p_T$ particles that generally do not significantly overlap with reference particles. Typically
these differential classes of particles have insufficient statistics from just POI angles to simply measure $v_n'$ using the same techniques as measurements of $v_n$ using charged particles.  Thus, the azimuthal anisotropies  that characterize the distributions of exclusively POIs are measured using "differential" correlators and cumulants which rely on correlations between POI and reference particles~\cite{Borghini:2000sa,Bilandzic:2010jr}.

Quantum-chromodynamics (QCD) and thereby the QGP, is required to locally conserve
quantities such as electric charge, baryon number, and strangeness.
Realistic handling of these quantities is necessary to accurately compare 
theoretical models to experimental data.
The study of fluctuations of $v_n^\prime$ for conserved quantities  is still in its infancy; however, recent work that includes baryon stopping in the initial conditions \cite{Shen:2017bsr} and other work that includes gluon splittings into quark anti-quark pairs \cite{Martinez:2019jbu,Carzon:2019qja} would allow one to study the fluctuations of conserved charges in the azimuthal anistropies; these studies may shed light on the effects of event-by-event fluctuations of baryon stopping and/or charge diffusion transport coefficients. 

Additionally, jets are of great interest in heavy ion collisions~\cite{Cunqueiro:2021wls}. They are created in large momentum transfer processes
in the very early stages of the collision prior to the fluid formation;
thus, they experience the same collision evolution as the fluid but they are
not equilibrated with it because the associated momentum scale is much
larger than the temperature of the fluid.
Jets are sensitive to the
short-length-scale properties of the QGP
and the average suppression of jets
in the QGP can be used to constrain the strength of jet 
quenching~\cite{JET:2013cls,He:2018gks,Ke:2020clc,JETSCAPE:2022jer}. 

In the case of jets, the value of $v_n^\prime$ is understood to be sensitive to the path-length dependence
of the interaction between the jets and the QGP~\cite{Wang:2000fq,Gyulassy:2000gk}
and measurements have been made which show positive values for these \vnprime\
quantities~\cite{ATLAS:2013ssy,ALICE:2015efi,ATLAS:2021ktw}.
Theoretically, hydrodynamical models have been used to elucidate decorrelations between the event plane angles $\Psi_n$ and $\Psi'_n$ for reference particles and for POI respectively, as well as for event plane angles between two different harmonics \cite{Heinz:2013bua,Qian:2016fpi,Qian:2017ier}. This decorrelation of POI event planes from the reference particle event plane is important when considering event-by-event fluctuations in $v'_n$. 
However, current techniques and observables do not provide a comprehensive way to study these jet-by-jet fluctuations 
in energy loss~\cite{Wicks:2005gt}.

 Measurements of mesons containing heavy, charm, and bottom quarks may provide interesting insight to various phenomena unique to heavy flavor particles.  At high momentum, heavy quarks come from jet production and suffer energy loss~\cite{Andronic:2015wma,Dong:2019byy}.  
At intermediate momenta, heavy quarks are understood to undergo Langevian like diffusion as they move through the QGP \cite{Moore:2004tg,Andronic:2015wma}.  Additionally, hadronization of heavy quarks is thought
to be modified in heavy ion collisions with recombination processes playing an important role~\cite{Cao:2013ita,Cao:2015hia,Katz:2019fkc}.
Due to all these effects, it is of great interest to measure the azimuthal anisotropies and their fluctuations 
of hadrons containing heavy quarks to constrain theoretical models~\cite{Nahrgang:2013saa,Sun:2019gxg,Prado:2016szr,Plumari:2019hzp}.
 
Finally, both jets and heavy flavor lead to ambiguous signals in proton-nucleus
collisions where  positive $v_n'$ values for high $p_T$ particles are measured~\cite{ATLAS:2019xqc}  but, there is no
significant  suppression~\cite{ATLAS:2014cpa,ALICE:2015umm,CMS:2016wma,ALICE:2021wct}.  Thus, studies of the fluctuations of $v_n^\prime$ could provide new information in these small systems to understand the origin of the observed $v_n^\prime$.

   In this work, we will focus on the development of  observables that utilize angular correlations between arbitrarily many reference particles with one or more POI. The POI selection varies with the physics of interest and could include, high $p_T$ jets, heavy-flavor hadrons, or some other object classification.
   Our explicit intent is to derive observables that are sensitive to not just the RMS of the $v_n$ fourier coefficients for POI, $\sqrt{\avg{{v_n^\prime}^2}}$ but also higher order fluctuations displayed by $v_n$.  
  
\begin{figure}
\centering
\includegraphics[width=0.85\textwidth]{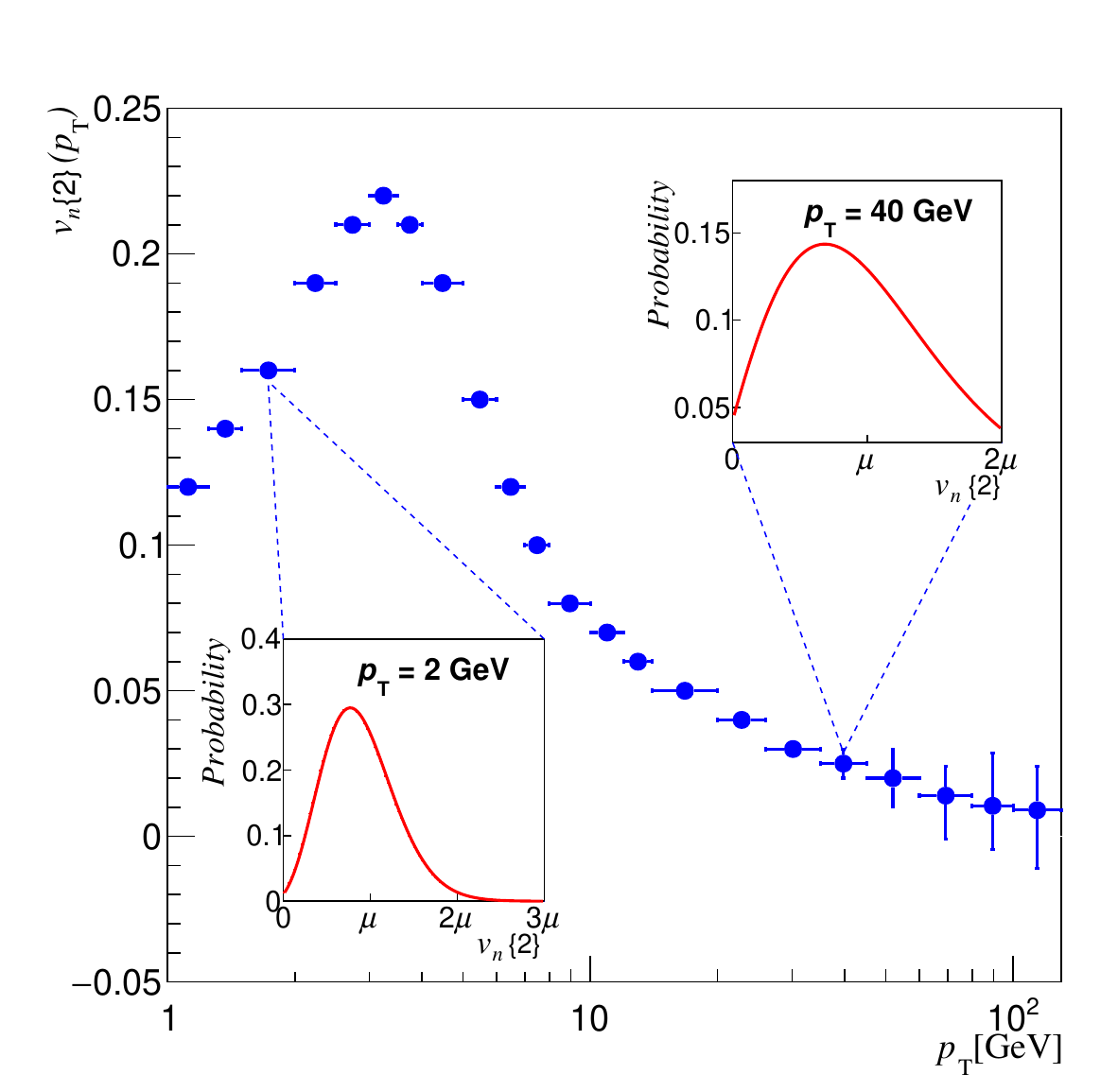}
\caption{Cartoon of $v_n\left\{2\right\}(p_T)$ as a function of \pt.  Insets
show examples of the distributions of $v_n\left\{2\right\}(p_T)$ for low \pt\
particles (dominated by reference particles) and 
high \pt\ (particles of interest). }
\label{fig:cartoon}
\end{figure}

   Generally, we expect the fluctuations of $v_n^\prime$ in POI to be affected by both the 
   initial geometry of the collision, as for the reference particles, 
   and by additional
   process specific fluctuations.    Measurements of
   these quantities could provide unique information about the mechanism of jet energy loss~\cite{Betz:2016ayq}. Therefore, we qualitatively expect
   the fluctuations in \vnprime\ to be larger than those observed in the soft sector. 
   This is illustrated in Fig.~\ref{fig:cartoon} which shows a typical example of differential 2 particle estimate for $v'_n:$ $v_2'\left\{2\right\}$
    as a function of \pt\ for hadrons. The plotted $v_2'\left\{2\right\}$ values are the average over the 
   events in that particular event selection, typically a centrality bin in heavy ion
   collisions.  The insets show example distributions of event-by-event \vtwo\ values
   relative to the mean.  The goal of this work is to suggest experimental observables 
   using multi-particle correlations to provide experimental access to these underlying fluctuation
   distributions.

Given the impact of these types of measurements in the soft sector, it is very likely that  fluctuations of \vnprime\  can provide new insight into the processes discussed above. 
   The very large data samples
   at the LHC~\cite{Citron:2018lsq}, the forthcoming upgrades at the LHC, and the very large data samples projected from sPHENIX~\cite{sPHENIXBUP2022}, provide new opportunities to make these measurements.  
Up until today, only a handful of experimental papers have explored multi-particle cumulants that contain one POI~\cite{CMS:2017xgk,CMS:2021qqk,ALICE:2022dtx,ALICE:2022zks,ATLAS:2015qwl,ALICE:2018gif}.
We expect this field to expand significantly with upcoming data.

In this paper we derive the formalism to study different types of multi-particle cumulants with an arbitrary number of POI in azimuthal anisotropy measurements for jets, and other rare probes. We generalize the following observables to include one or more POI:  differential cumulants developed in Refs.~\cite{Borghini:2001vi,Borghini:2000sa}, higher order moments of \vnprime, Multiharmonic Cumulants developed by Moravcova et al. \cite{Moravcova:2020wnf}, and Asymmetric Cumulants developed by Bilandzic et al \cite{Bilandzic:2021rgb}.
Finally, we propose central moments of arbitrary order in POI angle dependence, and two observables to estimate the contribution to a cumulant or moment for an arbitrary number of particles. Finally, we summarize our work by discussing features of each specific observable, and what types of fluctuations and POIs they are most ideal for measuring.

\section{Azimuthal Anisotropy and Correlators}

The event-by-event azimuthal anisotropies of particles in heavy ion collisions are studied using a Fourier expansion for the distribution of particle angles around the beam axis, where $v_n$ represents the contribution of the $n$-th harmonic to the net particle yield:
\begin{equation}\label{eq:v_n}
\frac{d N}{d \phi} \propto 1+2 \sum_{n=1}^{\infty} v_n \cos \left[n\left(\phi-\Psi_n\right)\right]
\end{equation}
where the detector angle $\phi_i$ for each of the $N$ particles is compared to the $n$-th order event plane; an azimuthal angle about which the distribution of $n\phi_i$ is symmetric \cite{Alver:2010gr}. When using the symbol  $v_n$, we specifically refer to the azimuthal anisotropy coefficients for reference particles, typically all measured charged particles. 

Likewise, we can define a related quantity, $v'_n$, as the Fourier coefficient for the distribution of POI angles. This quantity is an analogue to $v_n$, but generally relies on different approximation methods than $v_n$, due to properties of the probe in question:
\begin{equation}\label{eq:v'_n}
\frac{d N'}{d \psi} \propto 1+2 \sum_{n=1}^{\infty} v'_n \cos \left[n\left(\psi-\Psi'_n\right)\right]
\end{equation}
where the multiplicity of POI is labeled by $N'$ for a single event, and each POI is denoted with angle $\psi_i$ with respect to their theoretical event plane $\Psi'_n$, about which the angles $n\psi_i$ are symmetric. 

Using a Fourier expansion, we rewrite the explicit definition of both $v_n$ and $v'_n$ \cite{Borghini:2001vi,Borghini:2000sa}: 
\begin{equation}\label{eq:v_ndef}
     v_n \equiv\left\langle e^{i n\left(\phi-\Psi_n\right)}\right\rangle=\left\langle\cos [n\left(\phi-\Psi_n\right)]\right\rangle 
     \end{equation}
\begin{equation}\label{eq:v'_ndef}
     v'_n \equiv\left\langle e^{i n\left(\psi-\Psi'_n\right)}\right\rangle=\left\langle\cos[ n\left(\psi-\Psi'_n\right)]\right\rangle 
\end{equation}
where angle brackets $\avg{}$ indicate an average taken over all particle angles within a single event.  Flow harmonic vectors that contain information about $v_n$, and the $n$-th order event plane and can be defined as: 
\begin{eqnarray}\label{eq:vectors}
    V_n& \equiv &v_n e^{in\Psi_n} \\
    V_n^\prime & \equiv &v_n^\prime e^{in\Psi_n^\prime}
\end{eqnarray}
within a single event. The definition of these vectors allows for a more consise way to express the evaluation of $v_n$ and $v'_n$ to different powers using multi-particle angular correlations, as detailed in the next section. 

\subsection{Correlators}\label{sec:corrs}

Multi-particle correlations can determine $v_n$ and $v'
_n$ to different powers, a more accurate and computationally effective process than approximating the event plane, and measuring $v_n$ and $v'_n$ as they are defined in Eqs.~(\ref{eq:v_ndef},\ref{eq:v'_ndef}) \cite{Voloshin:2008dg,Poskanzer:1998yz}.
To measure correlations, we rely on the assumption that reference particles and POI are distributed symmetrically around $\Psi_n$ and $\Psi'_n$ respectively, which is accurate in the limits of $N,N' \gg 1$. If this assumption holds for the reference particles, it has been shown that we can correlate n-tuples of particle angles to approximate powers of $v_n$ for different harmonics \cite{Bilandzic:2013kga, Bhalerao:2011yg, Voloshin:2010ut}.
The correlation amongst reference particles can be written as a product of $v_n$ coefficients with an exponential dependence on event planes, or as a product of unconjugated and conjugated complex flow vectors $V_n, V^*_n$: 
\begin{eqnarray} \label{eq:k+m_ref}
\avg{k+m}_{n_1,...,n_{k}|n_{k+1},...,n_{m+k}} &\equiv& \left\langle e^{i\left(n_1 \phi_1+...+n_k \phi_k - n_{k+1}\phi_{k+1} - ... - n_{m+k}\phi_{m+k}\right)}\right\rangle \\
&=& v_{n_1} \cdots v_{n_m} e^{i\left(n_1 \Psi_{n_1}+...+n_k\Psi_{n_k} - n_{k+1}\Psi_{n_{k+1}} - ... -  n_{k+m}\Psi_{n_{k+m}}\right)}\\
&=&V_{n_1}\cdots V_{n_k} V_{n_{k+1}}^* \cdots V^*_{n_m}
\end{eqnarray}
for an arbitrary number of $k+m$ correlated particles. Each reference particle is labelled with angle $\phi_p$, numbered one through $N$, and the average on the RHS is taken over all $k+m$-tuples of particles. 
The indices $n_1,...,n_k$ indicate that the particles are "positively correlated;" their azimuthal angles have a positive sign in the exponential in the above equation. The indices $n_{k+1},...,n_{k+m}$ that follow the vertical bar $|$ indicate "negatively correlated" harmonics, which are subtracted in the above equation's exponential. We require the sum of the positively and negatively correlated harmonics to be zero: for any $\avg{k+m}$ we must have 
\begin{equation}\label{eqn:conditon_kkpmmp}
    \sum_{i=1}^k n_i = \sum_{j = 1}^{m} n_{k+j},
\end{equation}
otherwise the resulting quantity is isotropic, and will average to zero due to symmetry of particles around each event plane. 

To measure $v'_n$, and correlations between $v'_n$ and $v_n$ within an event, we apply the same assumption about event plane symmetry for POI, and incorporate their angles into correlations in the same way as in Eq.~(\ref{eq:k+m_ref}), generalizing to correlators that use $k'+m'$ POI and $k+m$ reference particles: 
\begin{equation}\label{eq:k+mfull}
\begin{aligned}
     &\avg{k'+m' + k+m}_{n_1,...,n_k,n'_{k+1},...n'_{k+k'}|n_{k+k'+1},...,n_{k+k'+m},n'_{k+k'+m+1}...n'_{k+k'+m+m'}} \\
     &\equiv \avg{\exp i\left[\sum_{\alpha=1}^{k} n_\alpha \phi_{\alpha} + \sum_{\beta = k+1}^{k+k'} n'_\beta \psi_{\beta} - \sum_{\gamma= k+k'+1}^{k+k'+m} n_\gamma \phi_{\gamma} - \sum_{\delta = k+k'+m+1}^{k+k'+m + m'} n'_\delta \psi_{\delta} \right]}
\end{aligned}
\end{equation}
where we have in particular $k'$ positively correlated POI harmonics, $m'$ negatively correlated POI harmonics, $k$ positively correlated reference particle harmonics, and $m$ negatively correlated reference particle harmonics. Each reference particle index iterates through all $N$ reference particles in the event with angle $\phi_p$, and likewise each of the $N'$ POI angles $\psi_p$ is iterated through by POI indices. 

Plugging Eqs.\ (\ref{eq:v_n},\ref{eq:v'_n}) into Eq.\ (\ref{eq:k+mfull}), we can convert the correlator with $k^\prime + m'$ arbitrary POI angles and $k+m$ reference particle angles into an approximation for $v_n$ and $v'_n$ at arbitrary power and harmonic number:
\begin{equation}\label{eq:k+mfull_vn}
   \begin{aligned}
\avg{k'+m'+k+m}_{n_1,...,n_k,n'_{k+1},...n'_{k+k'}|n_{k+k'+1},...,n_{k+k'+m},n'_{k+k'+m+1}...n'_{k+k'+m+m'}}\\
       = \prod_{\alpha = 1}^k v_{n_\alpha}e^{i n_\alpha \Psi_\alpha} \prod_{\beta = k}^{k+k'} (v'_{n_\beta})e^{i {n'_\beta} \Psi'_{\beta}}
       \prod_{\gamma = k+k'+1}^{k+k'+m} v_{n_\gamma}e^{-i n_\gamma \Psi_\gamma} \prod_{\delta = k+k'+m+1}^{k+k'+m+m'} (v'_{n_\delta})e^{-i {n'_\delta} \Psi'_{\delta}}\\
       = V_{n_1}\cdots V_{n_k}V'_{n_{k+1}}\cdots V'_{n_{k+k'}}V^*_{n_k+k'+1}\cdots V^*_{n_k+k'+m}V'^*_{n_{k+k'+m+1}}\cdots V'^*_{n_{k+k'+m'+m}}.
   \end{aligned} 
\end{equation}
Similar to the condition in Eq.\ (\ref{eqn:conditon_kkpmmp}), we require that the sum of harmonics produces an anisotropic quantity:
\begin{equation}\label{eqn:condition_wprime}
    \sum_{i = 1}^k n_{i} + \sum_{j = k+1}^{k+k'} n'_{j} = \sum_{g = k+k'+1}^{k+k'+m} n_{g} + \sum_{h = k+k'+m+1}^{k+k'+m+m'} n'_{h}
\end{equation}
for  Eq.\ (\ref{eq:k+mfull_vn}) to be used.
Otherwise, Eq.\ (\ref{eq:k+mfull_vn}) will yield values consistent with zero due to the symmetry planes. 

Evaluating $\avg{k'+m'+k+m}$ will generally produce some dependence on the angle between the event planes for $v_n$ and $v'_n$ at each of the different harmonics $n_i$, defined by the event plane angle in the exponential. When considering any two flow vectors $V_n$ and $V_m^*$, for either POI or reference particles, we can understand that this angle dependence is simply the dot product of the various vectors.
An example of a multi-harmonic correlation between $v'_3$ and the product of $v_2$ and $v_4$ is demonstrated below:
\begin{equation}\label{eq:ex1}
\begin{aligned}
    \avg{2'+2}_{2,4|3',3'} &=   v_2v_4(v'_3)^2 \cos{(2\Psi_2 + 4\Psi_4 - 6 \Psi'_3)} = \avg{e^{i(2\phi_1 + 4\phi_2 - 3\psi_3 - 3\psi_4)} } \\
    &= V_{2}V_{4}({V'}_{3}^*)^2
\end{aligned}
\end{equation}
where the harmonics $n_1 ... n_{4}$ obey the condition in Eq.\ (\ref{eqn:condition_wprime}).  Also, we remind the reader that the subscripts in $\phi_1 \dots \phi_4$ indicate the indices for the particles being correlated, not the harmonics.  

To evaluate Eq.\  (\ref{eq:k+m_ref}) using experimental data,  
an average is taken over all $k+m$-tuples with unique particle angles at each index -- indicated by a primed summation $\sum^{'}$ \cite{Voloshin:2010ut,Bilandzic:2013kga}:
\begin{equation}\label{primesum_ref}
    \avg{k+m} = \frac{(N - m - k)!}{N!}\sum^{\prime}_{1,...{k+m}} e^{i(n_1\phi_{1} +...+n_k\phi_{k}-n_{k+1}\phi_{k+1} - ... - n_{k+m}\phi_{k+m})}
\end{equation}
where the normalization factor $\frac{(N-k-m)!}{N!}$ is the reciprocal of the number of $k+m$ tuples of unique particles in an event with $N$ total particles. Likewise, in keeping with the evaluation of correlators using one POI index in Ref.~\cite{Voloshin:2010ut}, we can construct the same summations with arbitrary dependence on particles of interest, using $N$ reference particles and $N'$ POI by writing Eq.\ (\ref{eq:k+mfull}) as a primed summation: 
\begin{equation}\label{primesum_full}
\begin{aligned}
    \avg{k'+m'+k+m} =& \\ 
    \frac{(N - k- m)!}{N!}\frac{(N' - k' - m')!}{N'!}&\sum^{\prime} \exp i\left[\sum_{\alpha=1}^{k} n_\alpha \phi_{\alpha} + \sum_{\beta = k+1}^{k+k'} n'_\beta \psi_{\beta} - \sum_{\gamma= k+k'+1}^{k+k'+m} n_\gamma \phi_{\gamma} - \sum_{\delta = k+k'+m+1}^{k+k'+m + m'} n'_\delta \psi_{\delta} \right]
\end{aligned}
\end{equation}
where in this case we assume that there are no POI that are also considered reference particles, which we expect to be a reasonable definition in most analyses. For a more general treatment with non-negligible overlap 
between reference particles and POI, see Appendix \ref{ap:gen}.

Calculating a $k'+m'+k+m$ particle correlation by iterating over $k'+m'+k+m$-tuples can be computationally challenging for higher order correlations.  However, in Refs.~\cite{Bilandzic:2013kga,Bilandzic:2010jr} a much faster method to calculate these correlators using polynomials of the event Q$_n$ vector is introduced, which makes the calculations of these correlators much easier, but more complex to write analytically. Traditionally, a Q$_n$ vector is defined by averaging all of the particle angles within an event: 
\begin{equation}
    \text{Q}_{n} = \sum_{p = 0}^{N} e^{in\phi_p}
\end{equation}
where each reference particle has angle $\phi_p$ with respect to the event plane $\Psi_n$. In Appendix \ref{ap:Qvec} we introduce an algorithm to recursively write a correlation of the form $\avg{k'+m'+k+m}$ using Q$_n$ vectors, analogous to to the method described in Ref.~\cite{Bilandzic:2013kga}, but incorporating arbitrary POI dependence. 

While mathematically the correlators defined in this section evaluate products of $v_n$ as described in Eq.~(\ref{eq:k+mfull_vn}),  their values can be biased by sensitivity to nonflow fluctuations that contribute to $v_n$ in the form of correlations due to small subsets of particles  produced in heavy ion collision such as jets or resonance decays. These highly correlated particles fluctuate event-by-event, and do not necessarily reflect the collectivity induced by initial state geometry in the QGP. Additionally, because single event measurements fluctuate due to finite statistics, and the varying initial geometry of the QGP, it is more common to measure event by event averages of $v_n$ and associated quantities.
In the next section, we express the event-by-event averages of the correlations developed above, as the raw moments of a multivariate distribution over events, describe the stochastic variables that comprise the multivariate distribution, and relate them to existing observables.

\subsection{Raw Moments}\label{sec:rawmoments}

 Due to quantum mechanical fluctuations of the nucleons (or quarks and gluons) within the nucleus, as well as impact parameter fluctuations, there are fluctuating $v_n$ values on an event-by-event basis for a fixed centrality class \cite{ATLAS:2013xzf,Renk:2014jja,Shen:2015qta,Takahashi:2009na,Alver:2010gr,Alver:2010dn,Petersen:2010cw}.  The distribution of $v_n$ values for reference particles in a fixed centrality class has already been measured experimentally \cite{CMS:2013jlh}. The use of various underlying probability distributions to fit experimentally determined distributions for $v_n$ have been studied \cite{ATLAS:2013xzf}, and used to parametrize these experimental measurements, as well as simulation data, such as Gaussian \cite{Yi:2011hs}, Bessel Gaussian \cite{Voloshin:2008dg,ALICE:2022dtx}, Elliptic Gaussian, and Elliptic Power Distributions \cite{Yan:2013laa, Yan:2014afa}.  As of yet, no attempts have been made to experimentally measure the underlying distribution for $v_n^\prime$. Since a distribution and its fluctuations can be constrained by its moments or cumulants, we focus on methods to obtain moments and cumulants of $v_n^\prime$ to attempt to extract the underlying distribution.  

To better understand these distributions and their event-by-event fluctuations, we first consider the raw moments of $v_n$. Raw moments provide significant insight into various properties of the distribution's symmetry and tail behavior, and, more importantly, can always be used to express more conventional measures of fluctuations like central moments, or cumulants, which we will define later on in the paper.

The most natural way to define the raw moments in terms of the quantities already defined in the previous section would be to simply take event-by-event averages of $\avg{k'+m'+k+m}$. Since $\avg{k'+m'+k+m}$ measures a product of $V_{n_1}\cdots {V'}_{n_{m+k}}$, we understand the stochastic variables measured in these moments will also be a product of some collection of $r$ vectors: $V_{n_{i_1}}\cdots V'_{n_{i_r}}$. The exact determination of each stochastic variable represented in such a raw moment is addressed in  depth in the next section. Given an arbitrary collection of stochastic variables $X_1,...,X_n$, the raw moment, $\mu$, of order $\nu_i$ in each stochastic variable $X_i$ is defined as: 
\begin{equation}\label{eq:rawmoment_defn}
     \mu_{\nu_1, ...,\nu_m}(X_1 ... X_m) \equiv \left\langle \prod_{i=0}^n X_i^{\nu_i} \right\rangle
\end{equation}
where an average over all events is shown. 

 Earlier, we have defined multi-particle correlations, which specifically correlate $k+m$ reference particle angles at varying harmonics, and $k'+m'$ POI at varying harmonics in Eq.~(\ref{eq:k+mfull_vn}) \cite{Bilandzic:2013kga}.
  The expectation value of the correlator from Eq.~(\ref{eq:k+mfull_vn})  -- a raw moment -- is then obtained by averaging over an ensemble of events with index $i$, and weighting $W_i$, which is often uniform, based on total $p_T$, or total number of particles produced. This is indicated by the double brackets $\avg{\avg{}}$:
\begin{equation}\label{eq:averaging}
    \avg{\avg{k'+m'+k+m}} = \frac{\sum_{i}W_i \avg{k'+m'+k+m}_i}{\sum_{i}W_i}
\end{equation}
 and on the RHS, we also consider the \textit{single bracket} $\avg{}$ 
  to be a weighted average taken over events: 
\begin{equation}\label{eq:k+mavgfull}
   \begin{aligned}
\avg{\avg{k'+m'+k+m}}_{n_1,...,n_k,n'_{k+1},...n'_{k+k'}|n_{k+k'+1},...,n_{k+k'+m},n'_{k+k'+m+1}...n'_{k+k'+m+m'}}\\
       = \avg{\prod_{\alpha = 1}^k v_{n_\alpha}e^{i n_\alpha \Psi_\alpha} \prod_{\beta = k}^{k+k'} (v'_{n_\beta})e^{i {n'_\beta} \Psi'_{\beta}}
       \prod_{\gamma = k+k'+1}^{k+k'+m} v_{n_\gamma}e^{i n_\gamma \Psi_\gamma} \prod_{\delta = k+k'+m+1}^{k+k'+m+m'} (v'_{n_\delta})e^{i {n'_\delta} \Psi'_{\delta}}}\\
   \end{aligned} 
\end{equation}
where we have shown the full subindex of the correlator in Eq.\ (\ref{eq:k+mavgfull}) but suppressed it in Eq.\ (\ref{eq:averaging}) for simplicity's sake.

As a special case of this method, we note that correlators for exclusively reference particles, and correlators using one POI angle index have been defined explicitly in \cite{Borghini:2001vi, Bilandzic:2010jr}, using angle $\psi_p$ for POI angles.
The one POI differential correlator approximates the joint moment of order $2k-2$ in $v_n$ and first order in $V'_nV^*_n$. These correlators are used extensively in differential flow analyses with generating function cumulants to estimate values of $v'_n$ \cite{CMS:2017glf,CMS:2013wjq, ATLAS:2014qxy}: 
\begin{equation}
    \langle 1'+2k-1 \rangle = \langle e^{in(\psi_1 + \phi_1 .. \phi_{k-1} - \phi_{k} - ... - \phi_{2k-1})} \rangle = \frac{(N-2k+1)!}{N'(N!)} \sum^{\prime} e^{in(\psi_1 + \phi_1 .. \phi_{k-1} - \phi_{k} - ... - \phi_{2k-1})}
\end{equation}
where in the above equation we assume there is no overlap between POI and reference particles. After averaging this correlator over an ensemble of events, 
\begin{equation}\label{eq:1poi}
   \mu_{k-1,1}(v_n^2,V'_nV_n) = \langle\langle 1'+2k-1 \rangle\rangle 
\end{equation}
we can write it as a raw moment of the variabes $v_n^2$, and $V'_nV_n$.

As another example, we create a  raw moment from the example in Eq.~(\ref{eq:ex1}) by averaging the correlator $\avg{2'+2}_{2,4|3',3'}$:
\begin{equation}\label{eq:exampleavg}
     \avg{\avg{2'+2}}_{2,4|3',3'} = \avg{\left(v_2e^{i2\Psi_2}\right)\left(v_4 e^{i4\Psi_4}\right)\left(v'_3 e^{-i3\Psi'_3}\right)^2 } = \avg{v_2v_4(v'_3)^2 \cos{(2\Psi_2 + 4\Psi_4 - 6\Psi'_3)}}
\end{equation}
where we find that the condition in Eq.~(\ref{eqn:condition_wprime}) is satisfied just as before.  This raw moment identifies one of the contributions of $V_2$, $V_4$ and $V'_3$ to a multivariate distribution. We address how to determine the variables in the multivariate distribution, and address how to determine what correlations the above types of raw moments actually measured, in the next section. 

\subsection{Stochastic Variables and Normalization  } \label{vars}
 
Raw moments of a distribution often scale with the mean value of their distributions, $\langle v_n^m \rangle \sim \avg{v_n}^m$, making their comparisons difficult. For example, if stochastic variables $X$ and $Y$ are distributed normally with $\mu_2(X) = \mu_2(Y) = 5$, but $\mu_1(X) = 0.05$ and $\mu_1(Y) = 3$, the magnitude of their "fluctuations" relative to their mean are very different, despite both distributions having the same second raw moment. 
We define a normalized moment for a collection of stochastic variables $X_1,...,X_n$, denoted $N\mu_{\nu_1,...,\nu_n}(X_1,...,X_n)$: 
\begin{equation}\label{eq:rawNormalization}
    N\mu_{\nu_1,...,\nu_n}(X_1,...,X_n) = \frac{\mu_{\nu_1,...,\nu_n}(X_1,...,X_n)}{\prod_i^n \mu_1(X_i)^{\nu_i}}
\end{equation}
where the scaling of $\mu_{\nu_1,...,\nu_n}(X_1,...,X_n)$ that comes from $\avg{X_i}^{\nu_i}$ for each $i$ is cancelled by explicitly dividing by $\avg{X_i}^{\nu_i}$. If each stochastic variable $X_1,...,X_n$ is statistically independent of the others, then normalizing the raw moment will give unity; $N\mu_{\nu_1,...,\nu_n}(X_1,...,X_n)= 1$, and normalized moments that differ significantly from one indicate correlation or anticorrelation between the variables $X_1,...,X_n$. 

 Normalizing in this way is consistent with the existing normalization technique for Symmetric Cumulants (SC), which use the stochastic variables $X_1 = v_n^2$, and $X_2 = v_m^2$ to correlate the fourier coefficients of two harmonics \cite{Bilandzic:2013kga, Nasim:2016rfv, Mordasini:2019hut}. 
\begin{equation}\label{eq:SC}
    SC(v_n^2,v_m^2) = \avg{v_n^2v_m^2} - \avg{v_n^2}\avg{v_m^2}
    \end{equation}
    \begin{equation}\label{eq:NSC1}
    NSC(v_n^2,v_m^2) = \frac{\avg{v_n^2v_m^2} - \avg{v_n^2}\avg{v_m^2}}{\avg{v_n^2}\avg{v_m^2}}
\end{equation}
 It is clear that $SC$ is normalized by dividing $SC$ by each stochastic variable that it takes as an argument. It becomes intuitive to normalize raw moments of other stochastic variables the same way. We can now write $NSC(v_n^2,v_m^2)$ as a sum of a normalized raw moment and 1: 
\begin{equation}\label{eq:NSC2}
    NSC(v_n^2,v_m^2) = \frac{\avg{v_n^2v_m^2}}{\avg{v_n^2}\avg{v_m^2}} - 1 = N\mu_{1,1}(v_n^2,v_m^2) - 1
\end{equation}
 here, when considering quantities like $\avg{v_n^2,v_m^2} = \avg{\avg{2+2}}_{m,n|m,n}$, it is clear that $X_1 = v_n^2$ and $X_2 = v_m^2$ allows for consistency with the normalization for $SC$, as further discussed in Ref.~\cite{Bilandzic:2021rgb}. However, for quantities like the correlator defined in Eq.~(\ref{eq:ex1}), it is less clear how to select the correct stochastic variables. Additionally, for a correlator such as $\avg{\avg{2'+1}}_{2',2'|4}$ with odd numbers of particles, it's not well defined how to select the variables of interest in a manner consistent with Eqs.~(\ref{eq:SC}-\ref{eq:NSC2}). 

As shown in Eq.~(\ref{eq:k+mavgfull}), a raw moment of the form $\avg{\avg{k'+m'+k+m}}$ evaluates the event-by-event averages of products of flow vectors $V_n = v_ne^{in\Psi_n}$ and $V'_n=e^{in\Psi'_n}$. It is tempting to use these flow vectors as stochastic variables, but they cannot work within the normalization scheme, because $\mu_1(V_n)$ and $\mu_1(V'_n)$ are isotropic quantities and will average to zero, leaving the normalized raw moment in Eq.~(\ref{eq:rawNormalization}) undefined.

Instead, we use "nontrivial" stochastic variables that are selected by factoring out subsets of vectors for which the harmonics of vectors in  the subset add to zero.
If $\avg{k'+m' + k + m}$ evaluates to a product of flow vectors, 
\begin{equation}
\avg{k'+m' + k + m} = V_{n_1}...V_{n_k}V'_{n_{k+1}}...V'_{n_{k+k'}}V^*_{n_{k+k'+1}}V^*_{n_{k+k'+m}}V'^*_{n_{k+k'+m+1}}V'^*_{k+k'+m+m'}.
\end{equation}
we consider any disjoint subset of those vectors that produce a non-isotropic quantity to be a viable stochastic variable for our formalism. These groups can be 
represented by different correlators with fewer indices,
$X_i = \avg{k_i^{'} + m_i^{'} + k_i + m_i}$, and since they are disjoint, the product of each stochastic variable yields: $X_1\cdot...\cdot X_n = \avg{k'+m'+k+m}$. 
To validate that the correlators  $\avg{\avg{k'+m'+k+m}}$ are actually expectation values of the raw variables $X_i = \avg{k_i^{'} + m_i^{'} + k_i + m_i}$, we relate $\avg{\avg{k'+m'+k+m}}, X_i,$ and $\mu_{\nu_1,...,\nu_n}(X_1,...X_n)$ in the below equation:   
\begin{eqnarray}
    \avg{\avg{k'+m'+k+m}}
   &=& \avg{\prod_{i}\avg{k_i'+m_i'+k_i+m_i}}\\
    X_i &\equiv& \avg{k_i'+m_i'+k_i+m_i} \\
    \mu_{\nu_1,...,\nu_n}(X_1,...X_n) &=& \avg{\prod_{i} X_i^{\nu_i} } = \avg{\avg{k'+m'+k+m}}
\end{eqnarray}
where subscripts with each harmonic $_{n_1,n_2,...}$ are dropped for clarity. Clearly, $v_n^2$ and $v_m^2$ both remain consistent with this definition, since 
$\avg{2}_{n|n} = v_n^2$, and $\avg{2}_{m|m} = v_m^2$ are both isotropic quantities.

When only considering nontrivial stochastic variables, there exist many $\avg{\avg{k'+m'+k+m}}$ that cannot be further normalized if no smaller group of indices within $\avg{\avg{k'+m'+k+m}}$ can be isolated that will add to zero. Most notably, $\avg{\avg{2}}_{n|n}$ cannot be normalized further; it functions well as a stochastic variable, but it doesn't contain any "smaller" nontrivial stochastic variables.
Additionally, there exist correlators with more than one way to normalize, or select stochastic variables. The actual selection of stochastic variables is arbitrary, and can be changed for each individual analysis. For example, we can find two possible selections of stochastic  variables for  $\avg{\avg{2'+3}}_{2',2,1|3',2}$, although the raw moment will return the same value for each, the meaning of $\avg{\avg{2'+3}}$ will change based on normalization:
    \begin{equation}\label{eq:ex2}
        \frac{\avg{\avg{2'+3}}_{1,2',2|2,3'}}{\avg{\avg{1'+1}}_{2'|2}\avg{\avg{1'+2}}_{1,2|3'}} = N\mu_{1,1}\left((V'_2V^*_2),({V'}^*_3V_2V_1)\right) 
        \end{equation}
        \begin{equation}\label{eq:ex2.5}
        \frac{\avg{\avg{2'+3}}_{1,2',2|2,3'}}{\avg{\avg{2}}_{2|2}\avg{\avg{2'+1}}_{1,2'|3'}} = N\mu_{1,1}\left(v_2^2,({V'}^*_3V'_2V_1)\right) 
    \end{equation}
where we see that the raw moment is valid according to Eq.~(\ref{eqn:condition_wprime}), and likewise $2 = 2, 2+1=3$ ensuring that the stochastic variables are nontrivial. In the top equation, stochastic variables are selected to examine the correlation between $V'_2V^*_2$ and ${V'}^*_3V_2V_1$, which has greater sensitivity to the difference between $\Psi_2$ and $\Psi'_2$ than the quantities in the bottom equation: $v_2^2$, and $V'_3V'_2V_1$. 
A more intentional choice of stochastic variables and harmonics can prove useful for explicitly isolating correlation and decorrelation between event planes. This can be seen clearly when considering 4 particle correlators that share the same harmonic:
\begin{equation}\label{eq:ex3corr}
        \avg{\avg{2'+2}}_{2',2|2',2}= \mu_{1,1}({v'}_2^2, v_2^2 )
        \end{equation}
        \begin{equation}\label{eq:ex3.5corr}
        \avg{\avg{2'+2}}_{2',2'|2,2} = \mu_{2}(v'_2v_2\cos{4 (\Psi'_2 - \Psi_2}))
\end{equation}
where in the first case, the stochastic variables must be ${v_2}^2$ and ${v_2'}^2$ because in order to satisfy Eq.~(\ref{eqn:condition_wprime}) the POI and reference harmonics ($n',n$) add to 0 separately: $2'+ (-2') = 0 = 2 + (-2)$. Likewise, in the second case, the stochastic variable must be ${v'_2}v_2\cos 2
({\Psi'}_2-\Psi_2)$ because the only groups of harmonics that add to zero are $2' + (-2) = 0$, which is included twice. 
It's clear from the above equation that changing the position of $2$ and $2'$ relative to the bar allows us either to create a moment sensitive to the covariance in the magnitudes of $V_2$ and $V'_2$, or to create a moment sensitive to the variance of the dot product of these vectors $V'_2V_2^*$, a quantity that is strongly sensitive to $\Psi'_2 - \Psi_2$, the difference in angle between the differential event plane, and the reference particle event plane.\footnote{Exploiting this subtlety has given rise to the observable $A_n^f = \frac{\avg{v_n^2{v'}_n^2\cos2n({\Psi'_n - \Psi_n})}}{\avg{v_2^2{v'}_2^2}}$ in \cite{ALICE:2022dtx}. Generalizing this practice can be a useful alternative normalization scheme when trying to isolate decorrelations between reference and differential event planes.} 
Choices of this nature have been instrumental in isolating dependence on the angular flucutations between event plane angles for reference particles and POI, \cite{Heinz:2013bua,Qian:2017ier,Bilandzic:2021rgb,CMS:2017xnj,ATLAS:2017rij} or to exclude event plane dependence, and analyse exclusively the magnitude fluctuations between POI and reference $v_n$ and $v'_n$ coefficients \cite{ALICE:2022dtx}. 

\subsection{ Relation to Existing Observables}

Common observables measured in heavy ion collisions are two particle correlations used to estimate $v_n$ and $v'_n$. These can be easily recovered within the framework of POI correlators and raw moments defined in this section. The two particle correlation to measure $v_n$ is defined in Ref.~\cite{Borghini:2001vi} as:
\begin{equation}\label{eq:2ref}
    \left(v_n\{2\}\right)^2 = \avg{\avg{2}}_{n|n} = \avg{v_n^2}
\end{equation}
where $\{2\}$ indicates a two particle estimate for $v_n$. Likewise, a two particle estimate for $v'_n$ can be defined using a similar two particle correlation, but one reference particle index with a POI index:
\begin{equation}\label{eq:here}
        v'_n\{2\} = \frac{\avg{\avg{1'+1}}_{n'|n}}{\sqrt{\avg{\avg{2}}}_{n|n}} = \frac{\avg{V'_nV_n^*}}{\sqrt{\avg{v_n^2}}} \approx \avg{v'_n}  
\end{equation}
\begin{equation}
        v'_n\{2\}v_n\{2\} = \avg{\avg{1'+1}}_{n'|n} = \avg{V'_nV_n^*}
        \end{equation}
where the approximation in Eq.~(\ref{eq:here}) is valid in the limit that $\Psi'_n = \Psi_n$. 
More complicated $v'_n$ and $v_n$ cumulants require sums of more specific correlations, and will be used to motivate more general cumulants in Sec. \ref{gfc}. 

\section{Differential Moments and Cumulants}

\subsection{Fluctuations}

An important reason for studying fluctuations in $v_n$ and related observables in heavy ion collisions is to measure event-by-event fluctuations in the initial geometry of the collision \cite{Voloshin:2008dg}, and in the shape of the resulting QGP. Measuring the fluctuations of $v'_n$ for various probes, and relating them to fluctuations in $v_n$ will provide information about the sensitivity of these probes to the path length they travel through the QGP -- and how fluctuations in their path length produce fluctuations in their own abundance, angular distribution, and energy. 

If there were no statistical limitations, one could simply analyze the similarities in the probability distributions $p(v_n)$ and $p(v'_n)$. For the azimuthal anisotropy of reference particles, the distribution $p(v_n)$ has been studied, and compared to existing parametrizations for initial state anisotropy \cite{ATLAS:2013xzf}. In general, we expect the event-by-event distribution for differential azimuthal anisotropies, $p(v'_n)$ to be different from the event-by-event distribution of reference fourier coefficients $p(v_n)$: we expect both the means and relative fluctuations to be different between $p(v_n)$ and $p(v'_n)$. While there may not be sufficient particles of interest to approximate an event-by-event distribution for $p(v'_n)$, measuring raw moments will constrain the distribution $p(v'_n)$, and can also measure angle and magnitude correlations between $v'_n$ and $v_n$ \cite{ALICE:2022dtx}. Both of these correlations can be used to understand the path-length dependence of energy loss. 

While the raw moments as described above can help approximate correlations and constrain distributions, they can be combined into functions with more interpretive value. A measurement of the raw moment $\mu_{\nu_1,...,\nu_n}(X_1,...,X_n)$ still retains some dependence on smaller moments $\mu_{\tilde{\nu_1},...,\tilde{\nu_n}}(X_1,...,X_n)$, where $\tilde{\nu_i} \leq \nu_i$. To isolate the genuine contribution of order $\nu_i$ for each stochastic variables $X_i$ to a joint distribution $F(X_1,...,X_n)$, we use functions of the raw moments to construct cumulants according to the generating functions used in~\cite{Borghini:2001vi,Moravcova:2020wnf}, and the cumulant formalism introduced in Ref. ~\cite{Bilandzic:2021rgb}. Additionally, we introduce central moments that can be used to discern correlations between the "spread" or distance from the mean between different stochastic variables.

\subsection{Generating Function Cumulants}\label{gfc}

In Refs.~\cite{Borghini:2000sa,Borghini:2001vi}, 
 a set of cumulants for estimating $v_n$ and its fluctuations are defined using a unique generating function. Aside from the prevalence of these cumulants in literature, and well understood measurements \cite{CMS:2017xgk, CMS:2017xnj, ATLAS:2019peb}, a benefit of using these "generating function cumulants" is that they estimate the mean and fluctuations for various powers of $v_n$, and $v'_n$, while suppressing sensitivity to nonflow contributions. These cumulants are derived from a generating function following the formalism  outlined by Kubo in ~\cite{osti_4764483}, although they do not meet some of the convenient properties for cumulants specified in ~\cite{osti_4764483} (specifically the properties of Statistical Independence, Reduction, Semi-Invariance, and Homogeneity \cite{Bilandzic:2013kga, Bilandzic:2021rgb}). However, cumulants derived from these generating functions have been used effectively for the estimation of $v_n$, $v'_n$, and their fluctuations in experimental and theoretical contexts \cite{Voloshin:2010ut,Voloshin:2008dg,CMS:2017xgk}. Moreover, in Ref.~\cite{Bilandzic:2021rgb}, the authors recommend the continued study of generating function cumulants, under the understanding that these quantities are not true statistical cumulants according to the formalism introduced in Ref.~\cite{osti_4764483}.
The evaluation of generating functions using a higher number of POI angles as defined in this section can allow for an easy comparison with existing results for reference particles. Finally, the generating function cumulants defined here are the primary quantities which we discuss that can be interchanged to approximate $v'_n$ directly, or approximate correlations between $v'_n$ and $v_n$.

The cumulants of any distribution are defined as coefficients in the Taylor expansion of the multivariate moment generating function around $z = 0$, for some complex variable $z$. For azimuthal angles in heavy ion collisions, the generating function below \cite{DiFrancesco:2016srj} was used as a moment generating function to obtain cumulants:
\begin{equation}\label{eq:G}
G_n(z)= \avg{\prod_{j=1}^M\left(1+z^* e^{i n \phi_j}+z e^{-i n \phi_j}\right)}
\end{equation}
where $\phi_j$ corresponds to the $j$-th reference particle angle of $M$ reference particles produced in an event. 
Taking the event-by-event average $\avg{G_n(z)}$:
\begin{equation}\label{eq:avgG}
    \left\langle G_n(z)\right\rangle=\sum_{k=0}^{M / 2} \frac{|z|^{2 k}}{M^{2 k}}\left(\begin{array}{c}
M \\
k
\end{array}\right)\left(\begin{array}{c}
M-k \\
k
\end{array}\right)\avg{\left\langle e^{i n\left(\phi_1+\cdots+\phi_k-\phi_{k+1}-\cdots-\phi_{2 k}\right)}\right\rangle}
\end{equation}
generates every possible combination of correlations between reference particle angles of a given order when expanded. 

The cumulant generating function $\mathcal{C}_n(z)$ and differential cumulant generating function $\mathcal{D}_{p/n}(z)$ are defined in reference to $G_n(z)$ \cite{Borghini:2000sa,Borghini:2001vi}. $\mathcal{C}_n(z)$ is used to estimate univariate generating function cumulants for $v_n$ at different powers: 

\begin{equation}\label{eq:C}
    C_n(z) = \ln(\avg{G_n(z)})
\end{equation}
where, for fixed multiplicity $M$ in the limit of $M \gg 1$, we see that the above yields approximately $\mathcal{C}_n(z)\equiv M\left[\avg{\left\langle G_n(z)\right\rangle}^{1 / M}-1\right]$,  as described in ref.~\cite{Yan:2013laa,Borghini:2001vi}. The cumulant generating function approximating the natural log of the moment generating function is result fundamental to the definition of multivariate cumulants. In the differential regime, $\mathcal{D}_{p/n}(z)$ is used to estimate $v'_p$ for particles of interest:
\begin{equation}\label{eq:D}
    \mathcal{D}_{p / n}(z) \equiv \frac{\left\langle e^{i p \psi} G_n(z)\right\rangle}{\left\langle G_n(z)\right\rangle}
\end{equation}
where we see that $\mathcal{D}_{p / n}(z)$ represents the event-by-event correlation between POI angles and $G_n(z)$.
To calculate the cumulants, the generating functions can be decomposed into a power series of correlators and powers of $z$ and $z^*$, as shown below:
\begin{equation}\label{eq:Cpower}
    \mathcal{C}_n(z) = \sum_{k, m} \frac{z^{* k} z^m}{k ! m !}\avg{\avg{k+m}}_{\underbrace{n,...,n}_{k}|\underbrace{n,...,n}_{m}}
\end{equation}
and likewise, differential cumulants can be expressed as a power series in $z$ and $z^*$, with slightly different angular correlations:
\begin{equation}\label{eq:Dpower}
    \mathcal{D}_{p / n}(z) = \sum_{k, m} \frac{z^{* k} z^m}{k ! m !}\avg{\avg{1' + k+m}}_{p',\underbrace{n,...,n}_{k}|\underbrace{n,...,n}_{m}}
\end{equation}
where the averaged quantities $\avg{\avg{}}$ are simply raw moments defined in Eq.~(\ref{eq:k+mavgfull}).
Correlators where $k \neq m$ are isotropic, and average to zero for $\mathcal{C}_n(z)$, so we only consider the case when $m = k$. Matching orders in $|z|^2$ between Eq.~(\ref{eq:C}) and Eq.~(\ref{eq:Cpower}) gives the expression for the cumulant $c_n\{2k\}$ in terms of the correlators defined in the previous section. For example, we can see easily from Eq.~(\ref{eq:Cpower}) that the correlator of 1st order in $|z|^2$ is simply: 
\begin{equation}
    c_n\{2\} = \avg{\avg{e^{in(\phi_1-\phi_2)}}} = \avg{v_n^2}
\end{equation}
which is the two particle correlation from Eq.~(\ref{eq:2ref}). 
The representations of generating function cumulants for up to 14 particles ($k = m = 7$) can be found in Ref.~\cite{Moravcova:2020wnf}. 
Likewise, differential correlators where $m-k \neq p$ also average to zero. Again, by matching orders of $|z|^2$ between Eq.\ (\ref{eq:D}) and Eq.~(\ref{eq:Dpower}) it is possible to construct the "differential cumulants" $d_n\{m+k+1\}$.

Here, we propose cumulant generating functions for differential cumulants with two, and arbitrary POI dependence respectively. These generating functions produce cumulants with a dependence on angle tuples with a set number of POI angles, and an arbitrary number of reference particle angles. The two POI differential cumulant $\mathcal{F}_{p/n}(z)$ is defined similarly to $\mathcal{D}_{p/n}(z)$, but correlates $G_n(z)$ with a pair of POI angles,
\begin{equation}\label{eq:F}
\mathcal{F}_{p/n}(z) = \frac{\langle e^{ip(\psi_1 - \psi_2)}G_n(z) \rangle}{\langle G_n(z) \rangle}
\end{equation}
where $\psi_1,\psi_2$ are POI angles with index $1$ and $2$, similar to $\phi_1,\phi_2$ from Eq.~(\ref{eq:avgG}). 
For generating function cumulants that correlate arbitrary numbers of POI and reference angles both positively $k'$ and negatively $m'$, we can define a more general differential cumulant generating function:

\begin{equation}\label{eq:H}
\mathcal{H}^{k',m'}_{p/n}(z) = \frac{\langle e^{ip(\psi_1 + ... + \psi_k' - \psi_{k'+1} - ... - \psi_{k'+m'})} G_n(z) \rangle}{\langle G_n(z) \rangle}
\end{equation}
where $k'$ and $m'$ continue to indicate the number of separate positive and negative indices for POI angles.
Note that $\mathcal{F}_{p/n}(z)$ is just a special case of $\mathcal{H}^{1,1}_{p/n}(z)$. These multi-differential cumulants can once again be expanded in terms of joint differential moments, just as was seen for $\mathcal{D}_n(z)$ and $\mathcal{C}_n(z)$:
\begin{equation}\label{eq:FPower}
    \mathcal{F}_{p / n}(z) \equiv \sum_{k, m} \frac{z^{* k} z^m}{k ! m !}\avg{\avg{2'+k+m}}_{p',\underbrace{n,...,n}_{k}|p',\underbrace{n,...,n}_{m}}
\end{equation}
and:
\begin{equation}\label{eq:Hpower}
     \mathcal{H}_{p / n}^{k',m'}(z) \equiv \sum_{k, m} \frac{z^{* k} z^m}{k ! m !}\avg{\avg{k'+m'+k+m}}_{\underbrace{p',...,p'}_{k'},\underbrace{n,...,n}_{k}|\underbrace{p',...,p'}_{m'},\underbrace{n,...,n}_{m}}
\end{equation}
where we can see each raw moment used in the definitions of $\mathcal{F}_{p/n}(z)$ correlates pairs of POI angles, and the raw moments defining $\mathcal{H}_{p/n}^{k',m'}(z)$ rely on $k'+m'$ POI angles. 
By matching orders in $|z^2|$, just as was done for $\mathcal{C}_n(z)$ and $\mathcal{D}_{p/n}(z)$, we can obtain an expression for the cumulants $f_{p/n}\{k'+m'+k+m\}$ and $h^{k',m'}_{p/n}\{k'+m'+k+m\}$ generated by $\mathcal{F}_n(z)$ and $\mathcal{H}_{p/n}^{k',m'}(z)$, using correlators. Again, for $\mathcal{F}_n(z)$ we require that $k = m$. For $\mathcal{H}^{k',m'}_n(z)$ we require that $k-m = m'-k'$ to ensure that the quantity is not isotropic, because in general we will not have the same number of positively and negatively correlated POI angles. The first few orders are shown below for $\mathcal{F}_{n/n}(z)$,
\begin{equation}\label{eq:fn}
f_n\{2\}=  \left\langle\left\langle    2'\right\rangle\right\rangle_{n'|n'} 
\end{equation}
where, the two particle generating function cumulant is simply a two particle correlation using only POI. The higher orders are more complicated:
\begin{equation}
f_n\{4\}=  \left\langle\left\langle 2'+2 \right\rangle\right\rangle_{n,n'|n,n'} -\left\langle\left\langle 1'+1' \right\rangle\right\rangle_{n'|n'}\langle\langle 1+1\rangle\rangle_{n|n} \end{equation}
\text{and:}
\begin{equation}
\begin{aligned}
f_n\{6\} = \left\langle\left\langle 2'+4 \right\rangle\right\rangle&_{n',n,n,|n,n,n'}-4\left\langle\left\langle 2'+2 \right\rangle\right\rangle_{n',n|n,n'}\langle\langle 1+1\rangle\rangle_{n|n}+4\left\langle\left\langle 1'+1' \right\rangle\right\rangle_{n'|n'}\langle\langle 1+1\rangle\rangle_{n|n}^2\\
-&\left\langle\left\langle 1'+1' \right\rangle\right\rangle_{n'|n'}\langle\langle 2+2\rangle\rangle_{n,n|n,n}. \\
\end{aligned}
\end{equation}
We can see that in each equation, each term depends on a correlator with two POI angles. Also note that for each term, the number of particles correlated in each correlator sum to the order: for $f_n\{6\}$, each term has either a six particle correlator, a product of a two particle correlator and four particle correlator, or a product of three two particle correlators. This is consistent with existing cumulants for reference flow and differential flow using up to one POI \cite{Borghini:2001vi}.
Given $k'$ and $m'$ positive and negative separate indices for POI angles, we can evaluate $\mathcal{H}_{p/n}^{k',m'}(z)$ in much the same way, by matching terms in the power series expansion of Eqs. (\ref{eq:H}, \ref{eq:Hpower}). 

The coefficients found when evaluating $f_n\{k'+m'+k+m\}$ at different orders in Eq.~(\ref{eq:fn}) correspond exactly to coefficients in the Multivariate Harmonic Cumulants in Ref.~\cite{Moravcova:2020wnf}, which use generating function cumulants to correlate $v_n$ and $v_m$.
An area for further study would be to construct generating function cumulants to measure arbitrary powers of $v_n$ and $v'_n$ in different harmonics, using the full generality of the correlators defined in Eq.~(\ref{eq:k+mavgfull}).

 Now that the evaluation of generating function cumulants $f_{p/n}\{k'+m'+k+m\}$ or $h^{k',m'}_{p/n}\{k'+m'+k+m\}$ at each order in terms of raw moments is clear, we show how $f_{p/n}\{k'+m'+k+m\}$ and $h^{k',m'}_{p/n}\{k'+m'+k+m\}$ relate to $v_n$ and $v'_n$, and their fluctuations. Using the method in Refs.~\cite{Bilandzic:2012wva, Borghini:2001vi}, we calculate the contribution from $v_n$ and $v'_p$ to the values of the $f_n\{k'+m'+k+m\}$ cumulants. 
First, we define notation, in which $\left. \avg{x}\right|_{\Phi_n}$ indicates the average of  $x$ in all events with reaction plane angle $\Phi_n$. While the plane cannot be measured, we simply used it as a placeholder, varying between 0 and $2\pi$:
\begin{equation}\label{eq:solidangle}
        \left.\langle x\rangle \equiv \frac{1}{2 \pi} \int_0^{2 \pi/n}\langle x\rangle\right|_{\Phi_n} d \Phi_n
        \end{equation}
 indicating the average value of quantity $x$ at angle $\Phi_n$ before integrating around the entire transverse plane \cite{ALICE:2011ab}.
We now use this convention to express $v_n$ and $v'_{p/n}$ in the equation below \footnote{For this derivation, we use the  assumption that $\Psi_n \approx \Psi'_p \approx \Phi_n$: both event planes lie more or less on the reaction plane of the event. This assumption is reasonable for high multiplicity collisions and non-trivial $v_n$ and $v'_n$ values, and one required in the derivation for 1 POI differential flow cumulants. When this assumption fails, and event planes decorrelate, we will only measure the real part of  $v'_n$ projected onto $\Psi_n$}:   
        \begin{equation}
        \left.\left\langle e^{i n \phi}\right\rangle\right|_{\Phi_n}=\left.\left\langle e^{i n\left(\phi-\Phi_n\right)}\right\rangle\right|_{\Phi_n} e^{i n \Phi_n}=v_n e^{i n \Phi_n}
\end{equation}
\begin{equation}
    \left.\left\langle e^{i p \psi}\right\rangle\right|_{\Phi_n}=\left.\left\langle e^{i p\left(\psi-\Phi_n\right)}\right\rangle\right|_{\Phi_n} e^{i n \Phi_n}=v'_n e^{i n \Phi_n}
\end{equation}
where we consider integrating around the event plane to be approximately equivalent to an average taken over all particle angles. Using this notation, we examine $\avg{e^{ip(\psi_1 - \psi_2)}G_n(z)}$, and $\avg{G_n(z)}$ to evaluate $\mathcal{F}_{p/n}(z)$ with the same method as \cite{Bilandzic:2012wva}, by decomposing $\avg{}$ into an integral:
\begin{equation}
    \mathcal{F}_{p/n}(z) = \frac{\avg{e^{ip(\psi_1 - \psi_2)}G_n(z)}}{\avg{G_n(z)}}=\frac{\left.\frac{1}{2 \pi} \int_0^{2 \pi}\left\langle e^{i p\left(\psi_1-\psi_2\right)} G_n(z)\right\rangle\right|_{\Phi_n} d \Phi_n}{\avg{G_n(z)}}.
\end{equation}
We now use Eq.~(\ref{eq:solidangle}) to substitute in $\pm e^{ip\Psi_n}$, which immediately cancels, allowing us to remove $|v'_{p/n}|^2$ from the numerator. This shows that the power series expansion in $z^2$ of $\mathcal{F}_{p/n}$ approximates $|v'_{p/n}|^2$:
\begin{equation}
    \frac{\left.\frac{1}{2 \pi} \int_0^{2 \pi}\left\langle e^{i p\left(\psi_1-\psi_2\right)} G_n(z)\right\rangle\right|_{\Phi_n} d \Phi_n}{\avg{G_n(z)}} = \frac{\left.\frac{1}{2 \pi} \int_0^{2 \pi}\left\langle e^{i p\left(\psi_1 -\Phi_n + \Phi_n -\psi_2\right)} G_n(z)\right\rangle \right|_{\Phi_n} d \Phi_n}{\avg{G_n(z)}} 
\end{equation}
\begin{equation}
     \mathcal{F}_{p/n}(z)  =\frac{\left.\frac{v_{p / n}^{\prime} v_{p / n}^{\prime *}}{2 \pi} \int_0^{2 \pi}\left\langle G_n(z)\right\rangle\right|_{\Phi'_n} d \Phi'_n}{\avg{G_n(z)}}=\frac{\left|v_{p / n}^{\prime}\right|^2\left\langle G_n(z)\right\rangle}{\avg{G_n(z)}} = {v'_{p/n}}^2 \end{equation}
where the above displays consistency with cumulant estimates from $\mathcal{H}^{k',m'}_{p/n}(z)$, which will give values of $|{v'}_{p/n}|^{k'+m'}|v_n|^{k+m}$, with varying accuracy depending on the values of $k',m',k,m$ and the multiplicity of particles in the events being used. An explicit calculation of the contribution is performed in Appendix \ref{ap:C}.

Now that we have established $\mathcal{F}(z) = |v'_{p/n}|^2$, we can see it has no $z$ dependence, indicating that aside from  $f\{2\}$ (the 0th order $z$ contribution), the higher order cumulants $f\{2' + 2k\}$ \textit{do not scale with ${v'_n}^2$}, and instead measure corrections to $\mathcal{F}_{p/n}$ that come from the correlations between ${v'_n}^2$ and $v_n^2$. This is not surprising, when considering that $f_{p/n}\{4\}$ is a SC. While $f_{p/n}\{2'+2k\}$ does not scale with ${v'_n}^2$ beyond $f\{2\}$, different scaling behaviors can be selected with different values for $k',m',k$ and $m$ for $\mathcal{H}_{k,m}^{k',m'}$, which is further explained in Appendix \ref{ap:C}.

    \subsection{Symmetric and Asymmetric Cumulants}\label{sec:ASC}

Symmetric and Asymmetric Cumulants (ASC)  are statistical quantities used to correlate different even powers of $v_n$ and $v_m$, the azimuthal anisotropies of reference particles at different orders \cite{Bilandzic:2013kga, Bilandzic:2021rgb}. These quantities obey a number of fundamental properties as defined in~\cite{osti_4764483,Bilandzic:2021rgb}, and, thus, meet the rigorous mathematical definition of a cumulant for the variables $v_n^2$ and $v_m^2$. 
The simplest quantity of this form, the SC, is already described in Eq.~(\ref{eq:SC}). 

Cumulants of a set of stochastic variables $X_1,...,X_n$ isolate the "genuine" correlation between the variables, subtracting correlations between each subset of the stochastic variables, as well as the product of the variables. SC is widely used to measure the correlation between $v_n^2$ and $v_m^2$, and ASC extends the framework of SC to measure the correlations between larger sets of fourier harmonics $v_{n_1}^2 ...v_{n_l}^2$, or higher orders of dependence, measuring for example the correlations between $(v_n^2)^3$ and $v_m^2$.

We generalize the framework of ASC and SC to define cumulants for a selection of arbitrary variables $X_1,...,X_n$ as in Sec.~\ref{vars}, which each may contain arbitrarily many POI. We show these cumulants are consistent with the existing ASC and SC, and demonstrate how they can be used to intepret correlations and fluctuations in $v'_n$.

 A multivariate cumulant can be expressed as a set of derivatives of the cumulant generating function of a multivariate distribution with probability density $P(X_1,...,X_n)$:
    \begin{equation}\label{eq:cgf}
        C(\xi_1,...,\xi_n) = \ln\left({\int e^{(\xi_1X_1 + ... + \xi_nX_n)}P(X_1,...,X_n)dx_1...dx_n }\right)
    \end{equation}
where $C(\xi_1,...,\xi_n)$ is the generating function of a multivariate distribution, with dummy variables $\xi_1,...,\xi_n$, and stochastic variables $X_1,...,X_n$. Taylor expanding the cumulant generating function around $\xi_1 = ... = \xi_n = 0$ yields the cumulants:
    \begin{equation}
        \kappa_{\nu_1, \ldots, \nu_m}=\left.\frac{\partial^{\nu_1}}{\partial \xi_1^{\nu_1}} \cdots \frac{\partial^{\nu_m}}{\partial \xi_m^{\nu_m}} \ln \left\langle \exp\left({\sum_j \xi_j X_j}\right)\right\rangle\right|_{\xi_1=\xi_2=\cdots=\xi_m=0}
    \end{equation}
where the integral in Eq.~(\ref{eq:cgf}) is replaced with an expectation value.

Moreover, it is shown in Ref.~\cite{osti_4764483} that cumulants can be written as a function of raw moments as follows:    
\begin{equation}\label{eq:partition}
    \kappa_{1,...,1}\left(X_1, \ldots, X_m\right)=\sum^{P_m}_{l=1}(|l|-1) !(-1)^{|l|-1} \sum_{B\in P_m, |B| = l} \avg{\prod_{i \in B} X_i}
\end{equation}
where $P_m$ is the set of all partitions\footnote{A partition is a way to divide the a set of $m$ unique elements $\{1,...,m\}$ into subsets whose union contains the entire set $\{1,...,m\}$, but which does not use the same element twice. Then index $i$ is one "block" of the partition $B \in P_m$, a subset of $\{1,...,m\}$. For example, given the set $\{1,2,3,4\}$, we have $\{\{2,3\},\{1\},\{4\}\}$ and $\{\{1,3,4\},\{2\}\}$ are partitions, but $\{\{2\},\{2,3,4\}\}$ is not a partition because $1$ is not included in any subset, and $2$ is used twice. For the partition $\{\{1,3,4\},\{2\}\}$, the "blocks" $i \in B$ a partition are simply the elements: $\{2\}$ and $\{1,3,4\}$. 
The expression $|P_m|$ refers to the cardinality of $P_m$; the total number of partitions for the set $\{1,...,m\}$, while $|B|$ indicates the number of "blocks" in each partition $B \in P_m$.  The summation $\sum{B \in P_m, |B| = l}$ means that the summation is over all blocks in the partition which have $l$ elements.} of a subset of integers $\{1,...,m\}$. While it is hard to grasp intuitively from the above equation, the cumulant $k_{\nu_1, ...,\nu_n}(X_1,...,X_n)$ represents the subtraction of every "smaller" correlation between subsets of variables $X_{i_{1}}, ... X_{i_{\tilde{n}}}$ from the raw moment $\mu_{\nu_1, ...,\nu_n}(X_1,...,X_n)$.

Using Eq.~(\ref{eq:partition}), the first few cumulants can be calculated easily, although it becomes significantly more difficult at higher orders. We start with $\{1,2\}$: 
\begin{eqnarray}\label{eq:1,2}
        &P_2 = \Bigl\{\underbrace{\bigl\{\{1\},\{2\}\bigl\}}_{B_1}, \underbrace{\bigl\{1,2\bigl\}}_{B_2} \Bigl\}\\
        &\kappa_{1,1}(X,Y) = \avg{XY} - \avg{X}\avg{Y}
    \end{eqnarray}
and then form $\{1,2,3\}$:
\begin{eqnarray}\label{eq:1,2,3}
        &P_3 = \Bigl\{\underbrace{\bigl\{\{1\},\{2\},\{3\}\bigl\}}_{B_1}, \underbrace{\bigl\{1,2\bigl\}\bigl\{3\bigl\}}_{B_2},\underbrace{\bigl\{1,3\bigl\}\bigl\{2\bigl\}}_{B_3},\underbrace{\bigl\{2,3\bigl\}\bigl\{1\bigl\}}_{B_4},\underbrace{\bigl\{1,2,3\bigl\}}_{B_5} \Bigl\}\\
        &\kappa_{1,1,1}(X,Y,Z) = \avg{XYZ} - \avg{XY}\avg{Z}- \avg{XZ}\avg{Y} - \avg{ZY}\avg{X} + 2\avg{X}\avg{Y}\avg{Z}
\end{eqnarray}    
where $\kappa_{1,1}$ is an example of a SC and $\kappa_{1,1,1}$ is an example of a ASC.
In these equations we first compute the possible partitions that can be created with the sets $\{1,2\}$ and $\{1,2,3\}$. Then we evaluate the summation in Eq.~(\ref{eq:partition}).
The above relations can quickly be used to create cumulants of higher orders for specific variables because cumulants demonstrate reduction, allowing the variables in Eq.~(\ref{eq:partition}) to be  interchanged, and duplicated e.g.
 \begin{equation}\label{eq:multiv}
     \kappa_{1....1}(\underbrace{X, \ldots, X}_j, \underbrace{Y, \ldots, Y}_{k}) = \kappa_{j,k}(X,Y)
 \end{equation}
 as was  shown in \cite{Bilandzic:2021rgb, osti_4764483}.

While Eq.(\ref{eq:partition}) expresses how to define a cumulant with first order dependence on $n$ unique variables, we can see from this equation that if they are set equal $X_1 = X_2 = ... = X_n$, we will obtain an $n$-th order cumulant in $X_n$. This is demonstrated for the cumulant $k_{2}(X)$:  
\begin{equation}\label{eq:variance}
    \kappa_{1,1}(X,X) = \kappa_{2}(X) = \avg{X^2} - \avg{X}^2
\end{equation}
which can be compared to the result in Eq. (\ref{eq:1,2}). 
The coefficients obtained from the method for evaluating cumulants detailed in Eqs. (\ref{eq:partition} - \ref{eq:variance}) replicate the coefficients in the Asymmetric Cumulants defined in Ref.~\cite{Bilandzic:2021rgb}. We can simply replace $X,Y,Z$ with the same stochastic variables we could include in raw moments as defined in section \ref{vars}. It has been shown \cite{osti_4764483} that multivariate cumulants, like the ones defined here can always be written as a function of raw multivariate moments. Furthermore, we can see that $\kappa_{\nu_1,...,\nu_m}$ can be written as a function of raw moments of the same stochastic variables $\mu_{\nu_1,...,\nu_m}$. To demonstrate how each cumulant can be written as a sum of raw moments, we present some lower order bivariate cumulants of $v_n^2$ and ${v'}_n^2$, and a trivariate cumulant in Eqs. (\ref{eq:biv1}-\ref{eq:biv3}), and  Eq.~(\ref{eq:triv1}) respectively

\begin{equation}\label{eq:biv1}
\kappa_{1,1}(v_n^2, {v'}_n^{2})= \left\langle v_n^2 {v'}_n^2\right\rangle-\left\langle v_n^2\right\rangle\left\langle {v'}_n^2\right\rangle
\end{equation}
\begin{equation}\label{eq:biv2}
\kappa_{2,1}(v_n^2, {v'}_n^2)=  \left\langle v_n^4 {v'}_n^2\right\rangle-\left\langle v_n^4\right\rangle\left\langle {v'}_n^2\right\rangle-2\left\langle v_n^2 {v'}_n^2\right\rangle\left\langle v_n^2\right\rangle+2\left\langle v_n^2\right\rangle^2\left\langle {v'}_n^2\right\rangle \end{equation}
\begin{equation}\label{eq:biv3}
\begin{aligned}
\kappa_{3,1}(v_n^2, {v'}_n^2)=  &\left\langle v_n^6 {v'}_n^2\right\rangle-\left\langle v_n^6\right\rangle\left\langle {v'}_n^2\right\rangle-3\left\langle v_n^2 {v'}_n^2\right\rangle\left\langle v_n^4\right\rangle-3\left\langle v_n^4 {v'}_n^2\right\rangle\left\langle v_n^2\right\rangle +6\left\langle v_n^4\right\rangle\left\langle v_n^2\right\rangle\left\langle {v'}_n^2\right\rangle\\ &+6\left\langle v_n^2 {v'}_n^2\right\rangle\left\langle v_n^2\right\rangle^2-6\left\langle v_n^2\right\rangle^3\left\langle {v'}_n^2\right\rangle 
\end{aligned}
\end{equation}
\begin{equation}\label{eq:triv1}
\kappa_{1,1,1}(v_n^2, {v'}_n^2, v_m^2)= \left\langle v_n^2 v_m^2 {v'}_n^2\right\rangle-\left\langle v_n^2 v_m^2\right\rangle\left\langle {v'}_n^2\right\rangle-\left\langle v_m^2 {v'}_n^2\right\rangle\left\langle v_n^2\right\rangle - \left\langle v_n^2 {v'}_n^2\right\rangle\left\langle v_m^2\right\rangle+
2\left\langle v_n^2\right\rangle\left\langle v_m^2\right\rangle\left\langle {v'}_n^2\right\rangle 
\end{equation}
where, using Eq.(\ref{eq:k+mavgfull}), we can rewrite each average as expressed above $\avg{v_n^{2a} ({v'}_n)^{2b} v_m^{2c}} = \avg{\avg{2a'+2b+2c}}_{n',n,m|n',n,m}$. Unsurprisingly, the bivariate Asymmetric Cumulant of first order in $v_n^2$ and ${v'}_n^2$, $\kappa_{1,1}(v_n^2,{v'}_n^2),$ forms a Symmetric Cumulant, which has been well studied in Ref.~\cite{Bilandzic:2013kga}. 

Since cumulants can be written as sums of products of raw moments, they can also be be normalizated according to the scheme introduced in Sec.~\ref{vars}: 
\begin{equation}\label{eq:Ncumulant}
    N\kappa_{\nu_1,...,\nu_n}(X_1 ... X_n) = \frac{\kappa_{\nu_1,...,\nu_n}(X_1 ... X_n)}{\prod_i \avg{X_i}^\nu_i}
\end{equation}
and as a result, normalized cumulants $N\kappa_{\nu_1,...,\nu_m}$ can be written a sum of normalized moments $N\mu_{\nu_1,...,\nu_m}$. We show the normalization of $\kappa_{2,1}(v_n^2,{v'}_n^2)$ using the stochastic variables $\avg{v_n^2}$ and $\avg{v_n'^2}$, cancelling redundant terms in each fraction:
\begin{equation}
    N\kappa_{2,1}(v_n^2,{v'}_n^2) = \frac{\left\langle v_n^4 {v'}_n^2\right\rangle}{\avg{v_n^2}^2\avg{{v'}_n^2}} - \frac{\left\langle v_n^4\right\rangle}{\avg{v_n^2}^2} - 2\frac{\avg{v_n^2 {v'}_n^2}}{\avg{v_n^2}\avg{{v'}_n^2}} + 2
    \end{equation}
    \begin{equation}
    = N\mu_{2,1}(v_n^2,{v'}_n^2)  - N\mu_{2}(v_n^2) - 2N\mu_{1,1}(v_n^2,{v'_n}^2) + 2.
\end{equation}
where it can be seen that we obtain a function of normalized moments, up to a constant $+2$. 

\subsection{Central Moments}\label{sec:centralmoment}

Like cumulants, central moments also define measures of fluctuation of the distributions of $X_i$. Central moments in general do not obey reduction, a property of cumulants, but they are more straightforward to interpret: they correlate the distance from the mean ("spread") in a set of stochastic variables with the spread in each other stochastic variable. Their interpretation has been studied often, as kurtosis, variance, and skewness are all central moments which are often used to describe properties of probability distributions.

Given nontrivial stochastic variables $X_1,...,X_n$ as described in Sec.~\ref{vars}, requiring each $X_i$ to be some correlator $\avg{k'+m'+k+m}$ for which the harmonics that are correlated and anticorrelated cancel, the central moments for $X_1 ... X_n$, are defined as follows: 
\begin{equation}\label{eq:centralmoment}
\Tilde{\mu}_{\nu_1 ..., \nu_n}(X_1 ... X_n) \equiv \avg{\prod_{i = 1}^n (X_i - \avg{X_i})^{\nu_i} }.
\end{equation}
In general, these central moments can be expanded to functions of raw moments by using the multinomial theorem and the fact that $\avg{X+c} = \avg{X} + \avg{c}$. Applying the procedure to Eq.~(\ref{eq:centralmoment}) gives the following: 
\begin{equation}\label{eq:binomial}
        \Tilde{\mu}_{\nu_1,...,\nu_{n}} = \avg{(X_1 - \avg{X_1})^{\nu_1}\cdots (X_n-\avg{X_n})^{\nu_n}}
        \end{equation}
        \begin{equation}\label{eq:binomial2}
    = \sum_{s = 0}^{\nu}\left(\sum_{s_1+...+s_n = s} 
        \left(\begin{array}{c}
        s \\
        s_1,...,s_n
        \end{array}\right)\avg{X_1}^{s_1}\cdots \avg{X_n}^{s_n}\avg{X_1^{\nu_1-s_1}\cdots X_n^{\nu_n-s_n}}
        \right)
\end{equation}
where we define $\sum_i {\nu_i} = \nu$. The summation $\sum_{s_1+...+s_n = s}$ is taken over every combination of positive integers $s_1,...,s_n$ such that $s_1+...+s_n = s$, and $s_i \leq \nu_i$ for all $1\leq i \leq n$. The result in Eq.~(\ref{eq:binomial2}) is obtained by expanding $\prod_{i}(X_i - \avg{X_i})^{v_i}$ using multinomial theorem, before averaging the resulting quantity. In \ref{tab:2}, we explicitly write $\tilde{\mu}$ for $\nu = 2,3,$ and $4$, for arbitrary random variables $X,Y,Z,W$. 
Since Eq.~(\ref{eq:binomial2}) allows us to express $\Tilde{\mu}$ as a function of raw moments $\mu(X_i)$. Following the same methods used for raw moments and cumulants, we seek to evaluate correlations by normalizing these central moments. Since each central moment can be written as a function of raw moments, we can use the normalization scheme from Eq.~(\ref{eq:rawNormalization}), and normalize a central moment by dividing by the average of each stochastic variable $X_i$ that it describes\footnote{Note that conventionally, central moments are normalized not by the product of their stochastic variables, but by the standard deviation of their stochastic variables: $$N\tilde{\mu}_{\nu_1,...,\nu_n}(X_1,...,X_n)\equiv \frac{\mu_{\nu_1,...,\nu_n}(X_1,...,X_n)}{\prod_{i=1}^{n}\left(\sqrt{\avg{X_i^2}-\avg{X_i}^2}\right)^{\nu_i} }.$$ While this practice is well motivated statistically, we prefer to introduce normalization schemes that will ensure that even $\tilde{\mu}_{1,1}(v_n^2,v_m^2) = SC(v_n^2,v_m^2)$ can be normalized in a way that is consistent with existing normalizations for SC, and related variables. Additionally, the possibility of measuring a raw moment including stochastic variable $X_i: \mu_{...\nu_i...}(...X_i...)$ does not guarantee the capacity to accurately measure its standard deviation, a quantity that generally has greater dependence on each stochastic variable $\avg{X_i}^2$. }; 
\begin{equation} 
N\Tilde{\mu}_{\nu_1 ..., \nu_n}(X_1 ... X_n) = \frac{\Tilde{\mu}_{\nu_1 ..., \nu_n}(X_1 ... X_n)}{\prod_{i=1}^{n} \avg{X_i}^{\nu_i}}
\end{equation}
where the stochastic variables $X_i$ and their associated dependences $\nu_i$ are the same as in the definition of central moments in Eq.~(\ref{eq:centralmoment}). 

As an example of central moments, we express the multivariate central moments of order two in $V'_nV^*_n$, and order $r = 0,1,2$ in $v_n^2$:  
\begin{eqnarray}
     &\tilde{\mu}_{2,r}({V_n'}V_n^*, v_n^2) = \avg{(V_n'V_n^*- \langle V_n'V_n^* \rangle )^{2} (v_n^2- \langle v_n^2 \rangle )^{r}}\\
        &\begin{aligned}
        \tilde{\mu}_{2}(V_n'V_n^*) &= \avg{(V'_nV_n^*)} - \avg{V'_nV_n^*}^2\\
        \tilde{\mu}_{2,1}(V_n'V_n^*,v_n^2) &= \avg{(V'_nV_n^*)^2v_n^2} - \avg{v_n^2}\avg{(V'_nV_n^*)^2} - 2\avg{V'_nV_n^*}\avg{V'_nV_n^*v_n^2} + 2\avg{v_n^2}\avg{V'_nV_n^*}^2\\
        \tilde{\mu}_{2,2}({V_n'}V_n^*, v_n^2) &= \avg{(V'_nV_n^*)^2v_n^4} - 2\avg{v_n^2}\avg{(V'_nV_n^*)^2v_n^2} - 2\avg{V'_nV_n^*}\avg{(V'_nV_n^*)v_n^4} + \avg{v_n^2}^2\avg{(V'_nV_n^*)^2}\\
        & + 4\avg{V'_nV_n^*}\avg{v_n^2}\avg{(V'_nV_n^*)v_n^2} + \avg{V'_nV_n^*}^2\avg{v_n^4} - 3\avg{V'_nV_n^*}^2\avg{v_n^2}^2
    \end{aligned}
\end{eqnarray}
where the above quantities correspond to the variance of $V'_nV_n^*$ in the $r = 0$ case, the extent to which a deviation from the mean of $v_n^2$ is generally accompanied by a "squared" deviation from the mean of $V_n'V_n^*$, in the $r = 1$ case. Finally, in the $r=2$ case, $\tilde{\mu}(V'_nV_n^*,v_n^2)_{2,2}$ represents the extent to which a squared deviation from the mean of $v_n^2$ is accompanied by a squared deviation from the mean in $V'_nV_n^*$. 

While a measurement of a cumulant provides useful information about the various correlations between each stochastic variable, a measurement of $\tilde{\mu}$ by definition expresses how strongly a departure from the mean $\avg{X_i}$ is correlated to departures from the mean for other stochastic variables. This can be seen in the example above with $\tilde{\mu}_{2,1}$, and is a more traditional measure of fluctuation. While cumulants satisfy more mathematical properties, central moments also by definition include some  mathematical properties that we outline in Appendix \ref{ap:D}.
Additionally, in the univariate case, central moments like skewness and variance have already been used to study fluctuations in the distribution of $v_n$
\cite{Giacalone:2017uqx,Betz:2016ayq}.

\section{Functions of the moments ($\Gamma$, $\zeta$)}\label{sec:Gammazeta}

\subsection{Motivation for $\Gamma$, $\zeta$}

When studying fluctuations in differential azimuthal anisotropy coefficients $v'_n$, it can be helpful to make a comparison to reference azimuthal anisotropy fluctuations in $v_n$ to understand their relative magnitudes. 
Since each central moment, cumulant, and generating function cumulant defined in this paper can be represented as a sum of products of raw moments, the difference between any cumulants or central moments that require the same number of particles in their largest correlator can be rewritten as a function of differences between raw moments that require the same number of particles. 
Central moments, and cumulants with order $\nu \leq 3$ are unique in that they can always be expressed as a product of correlations and the averages of stochastic variables, as expressed in Eqs.~(\ref{eq:binomial}, \ref{eq:binomial2}). This means, when normalized, they can always be written exclusively as a sum of normalized moments and constants. The difference between any two central moments that require the same number of particles can likewise be decomposed into the differences of normalized raw moments, and possibly constants.

One example of this result comes from the comparison of two normalized six-particle central moments, of which one relies on six reference particles, and the other relies on four reference particles and two POI:
 
\begin{equation}\label{eq:skew}
        N\tilde{\mu}_{3}(v_n^2) = \frac{\avg{v_n^6}}{\avg{v_n^2}^3} -3\frac{\avg{v_n^4}}{\avg{v_n^2}^2} - 2
        \end{equation}
where the above equation is the normalized third central moment for $v_n^2$, and the below equation has dependence on $V'_nV_n^*$:
        \begin{equation}
        N\tilde{\mu}_{2,1}(V'_nV_n^*,v_n^2) = \frac{\avg{v_n^2(V'_nV_n^*)^2}}{\avg{v_n^2}\avg{V'_nV_n^*}^2} - 2\frac{\avg{v_n^2 V'_nV_n^*}}{\avg{v_n^2}\avg{V'_nV_n}} - \frac{\avg{(V'_nV_n^*)^2}}{\avg{V'_nV_n^*}^2} -2.
    \end{equation}
Clearly, both of the above central moments correlate the $n$ order harmonics of six particles within an event, but $N\tilde{\mu}_{2,1}(V'_nV_n^*,v_n^2)$ correlates deviations in $v_n^2$ with squared deviations in $V'_nV_n^*$,
whereas $N\tilde{\mu}_{3}(v_n^2)$ simply measures a quantity similar to the skewness of $v_n^2$. A direct subtraction of these quantities  yields the following:
    \begin{equation}
        N\tilde{\mu}_{3}(v_n^2) - N\tilde{\mu}_{2,1}(V'_nV_n^*,v_n^2) = \left( \frac{\avg{v_n^6}}{\avg{v_n^2}^3} - \frac{\avg{v_n^2(V'_nV_n^*)^2}}{\avg{v_n^2}\avg{V'_nV_n^*}^2}\right)  - 2\left(\frac{\avg{v_n^4}}{\avg{v_n^2}^2} - \frac{\avg{v_n^2(V'_nV_n^*)}}{\avg{v_n^2}\avg{V'_nV_n^*}}\right) - \left(\frac{\avg{v_n^4}}{\avg{v_n^2}^2} - \frac{\avg{(V'_nV_n^*)^2}}{\avg{V'_nV_n^*}^2}\right)
        \end{equation}
which we can then write as a difference in normalized \textit{raw} moments:    
        \begin{equation}\label{eq:normdiff}
        \begin{aligned}
         N\tilde{\mu}&_{3}(v_n^2) - N\tilde{\mu}_{2,1}(V'_nV_n^*,v_n^2) = \\
         &\left(N\mu_{3}(v_n^2)) - N\mu_{2,1}(V'_nV_n^*,v_n^2)\right) - 2\left( N\mu_{2}(v_n^2) - N\mu_{1,1}(V'_nV_n^*,v_n^2) \right) - \left(N\mu_{2}(v_n^2) - N\mu_{2}(V'_nV_n^*)\right).
        \end{aligned}
\end{equation}
We find that the difference in the normalized raw moments can easily be grouped by the number of particles required for each raw moment. Parenthesis in the above equation separate pairs of raw moments containing 6, 4, and 2 particles. 

Given two raw moments $\mu_{\nu_1,..,\nu_n}(X_1,...,X_n)$ and  $\mu_{\nu_1,..,\nu_n}(Y_1,...,Y_n)$ where $X_1,...,X_n$ and $Y_1,...,Y_n$ are two sets of nontrivial stochastic variables described in Sec.~\ref{vars}, with the same dependence (coefficient $\nu_i$) for each stochastic variable $X_i$ or $Y_i$, and differing dependence on $v'_n$, we can determine the correlations between $X_1,...,X_n$, and the correlations between $Y_1,...,Y_n$, as well as remove scaling with $X_i^{\nu_i}$ by simply normalizing the moments. To compare the normalized raw moments of $X_1,...,X_n$ with $Y_1,...,Y_n$, we introduce $\Gamma$ and $\zeta$:
\begin{equation}
    \Gamma_{\nu_1 ... \nu_n}(X_1, ... X_n;Y_1, ... Y_n) \equiv N\mu_{\nu_1 ... \nu_n}(X_1 ... X_n) - N\mu_{\nu_1 ... \nu_n}(Y_1 ... Y_n) 
\end{equation}
\begin{equation}
\zeta_{\nu_1 ... \nu_n}(X_1, ... X_n;Y_1, ... Y_n) \equiv \frac{N\mu_{\nu_1 ... \nu_n}(X_1 ... X_n)}{N\mu_{\nu_1 ... \nu_n}(Y_1 ... Y_n)}
\end{equation}
where $\Gamma$ is the \textit{difference} between normalized moments with the same dependence on $(X_1,...,X_n)$ and $(Y_1,...,Y_n)$, and $\zeta$ is the \textit{ratio} between normalized moments with the same dependence on $(X_1,...,X_n)$ and $(Y_1,...,Y_n)$. 
Since fluctuations in stochastic variables $X_1,...X_n$ and $Y_1,...,Y_n$ are measured by the normalized central moments, evaluating differences in central moments can describe the relative magnitude of fluctuations in $Y_1,...,Y_n$ relative to $X_1,...,X_n$. By decomposing the central moments using the methods above, we can determine how much of the difference in fluctuations between $Y_1,...,Y_n$ and $X_1,...,X_n$ comes from correlations involving a specific number of particles. In the example in Eqs.~(\ref{eq:skew}-\ref{eq:normdiff}) we can compare how two particle correlations, four particle correlations and six particle correlations all contribute to the difference between fluctuations isolated by $N\tilde{\mu}_{2,1}(V'_nV_n^*,v_n^2)$ and $N\tilde{\mu}_{3}(v_n^2)$.

    \subsection{ Applications of $\Gamma$ and $\zeta$ }

Already, there have been cases where $\Gamma$ and $\zeta$ quantities have been studied. In Ref.~\cite{Betz:2016ayq}, the quantity $\Delta^{sh}$ was derived, and is equivalent to 
$\Delta^{sh} = \Gamma_{1,1}(v_n^2,v_n^2;V'_nV_n^*,v_n^2)$ in our notation. It was used to evaluate decorrelation and between $v_n$ and ${v'}_n$ in hydrodynamical models to describe jet energy loss. This quantity can be expressed as: 
\begin{equation}
    \Gamma_{1,1}(v_n^2,v_n^2;V'_nV_n^*,v_n^2) = \frac{\avg{v_n^4}}{\avg{v_n^2}^2} - \frac{\avg{v_n^2V'_nV^*_n}}{\avg{v_n^2}\avg{V'_nV^*_n}}
\end{equation}
where we can see that a positive value for $\Gamma_{2;1,1}(v_n^2;V'_nV_n^*,v_n^2)$ obtained in this context indicates that the correlations in magnitude between $v_n^2$ and $V'_nV_n^*$ are not as great as the contribution of $v_n^4$ to the distribution $P(v_n^4)$, or essentially that $V'_nV^*_n$ displays suppressed fluctuations around its mean compared to $v_n^2$. In this instance, using $\Gamma$ to obtain a \textit{difference} between these quantities allowed the authors to establish an absolute scale to compare the fluctuations in $v'_n$ and $v_n$ for POI at different values of \pt, and using hydrodynamical models with different parameters.

Additionally, ALICE \cite{ALICE:2022dtx} developed the correlation $M_n^f$ that is equivalent to $\zeta_{1,1}(v_n^2,v_n^2;{v'}_n^2,v_n^2)$ in our notation. It was used to measure correlations between the magnitudes of $v_n(\pt)$ and $v_n$, and is expressed below:
Writing out $\zeta_{1,1}(v_n^2,v_n^2;{v'}_n^2,v_n^2)$, 
\begin{equation}
    \zeta_{1,1}(v_n^2,v_n^2;{v'_n}^2,v_n^2) = \frac{\avg{v_n^2{v'_n}^2}}{\avg{v_n^2}\avg{{v'_n}^2}}\frac{\avg{v_n^2}^2}{\avg{v_n^4}}
\end{equation}
we can see that $\zeta_{1,1}(v_n^2,v_n^2;{v'_n}^2,v_n^2)$ describes the magnitude of correlation between $v_n^2$ and ${v'_n}^2$ in reference to the magnitude of the correlation of $v_n^2$ with itself, comparing the two quantities with a ratio. Using a ratio in this instance allowed for a cleaner comparison of a wide variety of behaviors between theoretical models, as well as the value of the observable over large regions of centrality and \pt.

The authors of \cite{ALICE:2022dtx} also
developed the correlation $A_n^f$ that is equivalent to $A_n^f = \frac{\avg{\avg{2'+2}}_{n',n'|n,n}}{\avg{\avg{2'+2}}_{n',n|n',n}}$ in our notation. In a manner similar to that detailed in Eq. (\ref{eq:ex3corr}, \ref{eq:ex3.5corr}), $A_n^f$ allowed for them to monitor the extent to which the $\Psi'_2$ and $\Psi_2$ symmetry planes differ for different \pt\  selections of POI.

While the motivation for each of these correlations was not the same, they all fit into the formalism presented into this paper, and its associated interpretive context. By understanding $M_n^f$,  $A_n^f$, and $\Delta^{sh}$ as variations on the same quantity, we understand more from a comparison of their values. Likewise, we can understand each quantity as a difference between $NSC$, a normalized covariance, or alternately measuring the four particle contribution to arbitrary differences in central moments between $v_n^2$ and $V'_nV_n^*$ and ${v'}_n^2$ respectively.

In general, a measurement of $\Gamma$ corresponds more closely to the difference between central moments as detailed in Eq. (\ref{eq:normdiff}), while $\zeta$ as a ratio may be more experimentally feasible, providing a more stark numerical difference, with smaller error between the magnitudes of the raw moments it compares. Additionally $\zeta$ is useful for understanding the relative magnitude of two raw moments, whereas $\Gamma$ on its own is more useful for understanding the absolute difference in magnitude between two central moments. Regardless, both observables achieve the same purpose: a direct comparison between the relative fluctuations of $X_1,...,X_n$ and $Y_1,...,Y_n$ for a fixed number of particles.

\section{Discussion}

Each of the observables we have discussed in this paper has its own unique benefits, which may help to answer a broad class of questions related to measuring fluctuations in $v'_n$, correlations between powers of $v'_n$, and other azimuthal anisotropy measurements with rare probes. The study of each rare probe or identified particle imposes different statistical constraints, and motivations such that it is not guaranteed that observables that successfully measure fluctuations in heavy flavor quark $v'_n$ are suitable for the study of \pt\ dependent POI. To help discern between their features, we summarize all the observables we developed here into Table~\ref{tab:1} that discusses their potential uses as well as caveats due to available statistics.

\begin{table}
\centering
\caption{ A summary of observable quantities defined in this paper, and their prospects for measuring fluctuations in $v'_n$ in heavy ion collisions. }
\begin{tabular}{p{0.35\linewidth}  p{0.65\linewidth}}
\hline
\hline
\multicolumn{1}{l}{\parbox{0.3\linewidth}{\centering \textbf{Observables}}} &  \multicolumn{1}{c}{\parbox{0.3\linewidth}{\centering \textbf{Explanation}}}\\

  \multicolumn{1}{l}{\parbox{0.3\linewidth}{\centering \textbf{Correlators}}} &  \multirow{2}{*}{\parbox{\linewidth}{\centering }}\\
  \multicolumn{1}{l}{\parbox{0.3\linewidth}{$\avg{k'+m'+k+m}
  $}}\\
  \multicolumn{1}{l}{\parbox{0.3\linewidth}{\centering }} &  \multirow{2}{*}{\parbox{\linewidth}{\centering  These correlators evaluate multiharmonic products of event flow vectors with arbitrary dependence on POI angles. They can be calculated using Q$_n$ vectors.}}\\
  \multicolumn{1}{l}{\parbox{0.3\linewidth}}\\
  
   \multicolumn{1}{l}{\parbox{0.3\linewidth}{\centering \textbf{Raw Moments }}} &  \\ 
   \multicolumn{1}{l}{\parbox{0.3\linewidth}{}} &  \multirow{4}{*}{\parbox{\linewidth}{\centering 
   Raw moments evaluate the expectation value of a product of flow vectors
   $\avg{V_{n_1}\cdots V_{n_k+m}V'_{n_{k+m+1}}\cdots V'_{n_{k'+m'+m+k}}}$ by taking a weighted average over many events. Selection of stochastic variables $X_i$ can be done by considering smaller groups of $V_{{n}_{1_i}}\cdots V'_{{n_i}_{k_i}}$ for which $\sum_{j} n_{j_i} = 0$. This selection of stochastic variables $X_i$ allows for a normalization scheme. }} \\ 
   \multicolumn{1}{l}{ \parbox{0.3\linewidth}{$\mu_{\nu_1,...,\nu_n}(X_1,...,X_n)$}} \\
   \multicolumn{1}{l}{ \parbox{0.3\linewidth}{ $\avg{\avg{k'+m'+k+m}}$}}\\
   \multicolumn{1}{l}{}\\
   \multicolumn{1}{l}{}\\
    \multicolumn{1}{l}{}\\
  
  \multicolumn{1}{l}{\parbox{0.3\linewidth}{\centering \textbf{Generating Function Cumulant}}} &  \multirow{2}{*}{\parbox{\linewidth}{\centering   }} \\
  \multicolumn{1}{l}{ $f_n\{k'+m'+k+m\}, h_n^{k',m'}\{k'+m'+k+m\}$} & \multirow{4}{*}{\parbox{\linewidth}{\centering Generating function cumulants use $\avg{\avg{k'+m'+k+m}}$ to approximate ${v'}_n^{k'+m'}$ using generating Functions. These observables are analogues of $c_n\{2k\}$ and $d_{p/n}\{2k\}$, with higher dependence on POI.}} \\
   \multicolumn{1}{l}{}\\
   \multicolumn{1}{l}{}\\
   \multicolumn{1}{l}{}\\
  
  \multicolumn{1}{l}{\parbox{0.3\linewidth}{\centering \textbf{Multivariate Cumulants}}}  & \multirow{5}{*}{\parbox{\linewidth}{\centering }} \\
   \multicolumn{1}{l}{\parbox{0.3\linewidth}{ $\kappa_{\nu_1,...,\nu_n}(X_1,...,X_n)$}}  & \multirow{5}{*}{\parbox{\linewidth}{\centering A cumulant of arbitrary order in each variable. Multivariate Cumulants generalize the framework of multiharmonic cumulants introduced in Ref.~\cite{Bilandzic:2021rgb} to a larger set of stochastic variables and POI. These observables in general obey more statistical properties than generating function cumulants, and represent the genuine contribution of a moment at each order to fluctuations, by subtracting autocorrelations.}} \\
   \multicolumn{1}{l}{}\\
   \multicolumn{1}{l}{}\\
  \multicolumn{1}{l}{}\\
   \multicolumn{1}{l}{}\\
   \multicolumn{1}{l}{}\\
  
  \multicolumn{1}{l}{\parbox{0.3\linewidth}{\centering \textbf{Multivariate Central Moment}}} & \multirow{3}{*}{\parbox{\linewidth}{\centering }} \\ 
   \multicolumn{1}{l}{\parbox{0.3\linewidth}{$\tilde{\mu}_{\nu_1,...,\nu_n}(X_1,...,X_n)$}} & \multirow{3}{*}{\parbox{\linewidth}{\centering Central moments explicitly describe the correlation between many different variables, and are traditionally used to calculate the higher order fluctuations in a distribution. They directly measure correlations of spread around the mean between stochastic variables.}} \\ 
\multicolumn{1}{l}{}\\
  \multicolumn{1}{l}{}\\
   \multicolumn{1}{l}{}\\

  \multicolumn{1}{l}{\parbox{0.3\linewidth}{\centering \textbf{Differences of Normalized Moments }}}  &  \multirow{3}{*}{\parbox{\linewidth}{}} \\ 
  \\
  \multicolumn{1}{l}{\parbox{0.3\linewidth}{ $\Gamma_{\nu_1,...,\nu_n}(X_1,...,X_n; Y_1,...Y_n)$}}  &  \multirow{3}{*}{\parbox{\linewidth}{\centering The differences between two normalized raw moments can be used to decompose the difference between two central moments of the same order, but in different variables $X_i$ and $Y_i$. Using $\Gamma$, we can discern the contribution at each order to a difference between central moments.}} \\ 
\multicolumn{1}{l}{}\\
  \multicolumn{1}{l}{}\\
   \multicolumn{1}{l}{}\\
  
  \multicolumn{1}{l}{\parbox{0.3\linewidth}{\centering  \textbf{Ratios of normalized moments}}} &  \multirow{3}{*}{\parbox{\linewidth}{\centering}} \\ 
  \multicolumn{1}{l}{\parbox{0.3\linewidth}{$\zeta_{\nu_1,...,\nu_n}(X_1,...,X_n; Y_1,...Y_n)$}} &  \multirow{3}{*}{\parbox{\linewidth}{\centering Taking the ratio of two normalized raw moments of different variables  at the same order allows a better understanding of the relative correlations of each raw moment and its constituent stochastic variables. These quantities are analogous to $\Gamma$, but may be easier to experimentally determine because they consitute a ratio.}} \\ 
\multicolumn{1}{l}{}\\
  \multicolumn{1}{l}{}\\
   \multicolumn{1}{l}{}\\
   \multicolumn{1}{l}{}\\
  
  \hline
  \hline
\end{tabular}
\label{tab:1}
\end{table}

First, $\avg{k'+m'+k+m}$, as detailed in Sec.~\ref{sec:corrs} can be used in a given event to evaluate the products of $V_n$ and $V'_n$ to different powers, and at different harmonics, by using correlations of azimuthal angles between harmonics. These estimations for the products of $V_n$ and $V'_n$ provide the basis for evaluating event-by-event fluctuations in $V_n$ and $V'_n$, so they are averaged over events to evaluate raw moments.

The joint raw moments with dependence on $v'_n$, $\avg{\avg{k'+m'+k+m}}$ are necessary to calculate any of the more complex observables described in this paper. In general, a measurement of these quantities, as shown in Sec.~\ref{sec:rawmoments} will provide enough information to additionally calculate and normalize any of the Generating Function Cumulants, Asymmetric Cumulants, central moments, and by extension $\Gamma$ and $\zeta$. However, each one of the above observables requires values for different $\avg{\avg{k'+m'+k+m}}$ moments; a choice of a more complex observable will determine which joint raw moments must be measured. 

Generating Function Cumulants, defined in Sec.~\ref{gfc} are unique from the other observables introduced here in that they can both estimate $v'_n$ to different powers, or evaluate correlations between $v'_n$ and $v_n$ using different orders of dependence on azimuthal angles 
from POI. Naturally, these quantities are suitable for comparison to existing generating function cumulant measurements of reference $v_n$ using the same number of reference particles $v_n\{2k\}$, and differential cumulant estimates for $v_n$ using only one particle of interest. Understanding how the contribution from the correlation of two or more POI azimuthal angles to a measurement of $v'_n$ differs from the contribution of one POI will provide unique and novel information about the azimuthal anisotropy of POI.

Asymmetric cumulants do not directly estimate $v'_n$ or any differential fourier harmonic. However, as shown in Sec.~\ref{sec:ASC} they can isolate the genuine correlations between $v'_n$ with $v_n$, and compare to experimental results, which have recently been studied in a multiharmonic context \cite{ALICE:2023lwx}. Additionally, since the cumulants $k_\nu(X), \;\nu >2$ of a Gaussian distribution are always 0, asymmetric cumulants can be used to evaluate "deviations from Gaussianity" in univariate distributions, \cite{CMS:2017glf}. 

The central moments we have introduced in Sec.~\ref{sec:centralmoment} bear many similarities to the asymmetric cumulants, and are identical at low order: $\kappa_{\nu_1,...,\nu_n}(X_1,...,X_n) = \tilde\mu_{\nu_1,...,\nu_n}(X_1,...,X_n),$ for $ \sum_i \nu_i \leq 3$. Measuring a central moment is useful for comparisons with skewness measured in Ref.~\cite{Giacalone:2017uqx,Giacalone:2016eyu} or variance defined in Refs.~\cite{Betz:2016ayq,Voloshin:2008dg,CMS:2013jlh}. Additionally, the measurement of a central moment provides knowledge about the correlations between $v_n$ and $v'_n$, and their dispersion around their means to different powers. 

In Sec.~\ref{sec:Gammazeta}, we showed the difference between the normalized raw moments of two collections of variables, $\Gamma_{\nu_1,...,\nu_n}(X_1,....,X_n;Y_1,...Y_n)$, is more useful for determining the relative contribution at each order from $v'_n$ to a large multivariate distribution of $v_n$ and $v'_n$ at different powers and harmonics. Additionally, we have shown that the difference between any two normalized central moments, including the traditional Symmetric Cumulant can be written as a linear combination of $\Gamma$. This means $\Gamma$ can determine the differences in correlations between stochastic variables including $v'_n$ and stochastic variables relying only on $v_n$ as a way of comparing the fluctuations in $v'_n$ and $v_n$.

While $\zeta$, also defined in Sec.~\ref{sec:Gammazeta}, cannot decompose the differences between  central moments or cumulants, (since it is a ratio rather than a difference) $\zeta$ is perhaps more easily measurable experimentally, as it is normalized to one, whereas $\Gamma$ can be both negative and positive. Additionally, $\zeta_{\nu_1,...,\nu_n}(X_1,..,X_n;Y_1,...,Y_n)$ can identify the \textit{relative magnitude} of fluctuations in $Y_1,...,Y_n$ versus fluctuations in $X_1,...,X_n$, whereas $\Gamma_{\nu_1,...,\nu_n}(X_1,..,X_n;Y_1,...,Y_n)$ can only describe their absolute differences. 

We anticipate the use of these variables to explore differential phenomena like the event-by-event fluctuations in energy loss for jets traversing the QGP medium
and other related questions. By evaluating $\zeta$ for sets of stochastic variables involving the azimuthal anisotropy of jets, we can compare the fluctuations of jet energy loss - probed by fluctuations and correlations between the suppression of $v'_2$ and $v_2$.

\section{Conclusion}

In this paper we introduce a way to apply existing statistical correlations and observables of azimuthal anisotropies to the study of rare probes and identified particles in heavy ion collisions.  To achieve this goal, we allowed for the inclusion of arbitrarily many unique particle of interest (POI) indices in differential correlators to provide an arbitrary dependence on differential azimuthal anisotropies. We lay out the methods to use these  multi-POI differential correlators to construct raw and central moments of their underlying distribution(s), generating functions for cumulants, asymmetric cumulants, and ratios or differences of correlations. We also outline various methods for properly normalizing these new observables. Examples are also provided for certain harmonics and POI dependence to guide the reader on how to construct these new observables.

The purpose of these observables is to study rare probes such as jets and/or heavy mesons as well as identified particles like strange hadrons.  These new observables will provide a method to constrain the fluctuations in the underlying distribution that is only now possible experimentally in the high luminosity era of the LHC and sPHENIX~\cite{Citron:2018lsq,Achenbach:2023pba,Arslandok:2023utm,sPHENIXBUP2022}. 
In another work, we studied \cite{pheno_paper} the feasibility of extracting different type of underlying distributions with these new observables, depending on the type of distribution and the magnitude of the observable calculated. While future studies are warranted to determine the statistics required to obtain precise measurements of these observables, we anticipate that experimental measurements and theoretical calcuations of these observables will prove useful in understanding properties of the QGP with identified particles.

\section{Acknowledgements }

The authors thank Matt Luzum and Jean-Yves Ollitrault
for fruitful and clarifying conversations that helped to construct the formalism presented here. A.H., A.M.S., and X. W. acknowledge support from  National Science Foundation
Award Number 2111046.
J.N.H. acknowledges the support from
the US-DOE Nuclear Science Grant No. DESC0020633 and DE-SC0023861 and and within the framework of the Saturated Glue (SURGE)
Topical Theory Collaboration.

\bibliographystyle{unsrturl}
\bibliography{refs,refs_noinspire}

\appendix
\section{Non-trivial overlap between POI and Reference particles}\label{ap:gen}

In the analysis of the azimuthal anisotropies of jets and high \pt\ particles, the high \pt\ POI are rarely included in the set of reference particles. Likewise for the azimuthal anisotropies of charm and bottom mesons, the overlap between charm mesons and reference particles is suppressed, simply due to the scarcity of these mesons. However, in general, there are some cases in which it is plausible that
scenarios with overlap between POI and reference particles are of interest. In this instance, the actual computation of $\avg{k'+m
+k+m}$ becomes somewhat more complex. In this appendix we discuss these cases. 

Using a summation over all $m'+k'+m+k$ tuples consisting of $m+k$ reference particle angles, and $m'+k'$ POI angles, we can see that considering the existence of $N_q$ overlapping POI and reference particles, we obtain a different weighting for the value of $\avg{k'+m'+k+m}$ than seen in Eq.~(\ref{primesum_full}). The summation for a correlator weighs each $m+k$-tuple of POI and reference particle angles the same, but only considers tuples in which each angle is distinct from the others, to omit autocorrelations. To calculate the average, we simply complete the primed summation, and divide by the number of $m+k$ tuples in which each angle is unique. Since there are $N$ reference particles, $N'$ POI and $N_q$ reference particles that are also considered to be POI, we can evaluate: 
\begin{enumerate}
\item for $k+m$ reference particle indices, the number of unique tuples is simply the number of ways to arrange $N$ particles in $k+m$ slots: $\frac{N!}{(N-m-k)!}$. 
\item Having selected $k+m$ reference particles, we fill $k'+m'$ more slots with POI. If there is no overlap between reference particles and POI, we receive the same result as before; the number of tuples is now $\frac{N'!}{(N'-k'-m')!}$. If there is $N_q$ overlap between POI and reference particles, then the number of choices is reduced: the computation in step (1) is the same, but now there are only $N' - (k+m)N_q$ POI to "choose" from. Like always, they are placed into $k'+m'$ slots, and, thus, we have $\frac{N'!-(k+m)N_q}{(N'-(k+m)N_q - k'-m')!}$ choices. 
\end{enumerate}
Since the total number of unique n-tuples is multiplicative, we obtain the following equation to calculate the correlator assuming a uniform weighting for each particle:
\begin{equation}\label{eq:totalmult}
   \begin{aligned}
\avg{k'+m'+k+m}_{n_1,...,n_k,n'_{k+1},...n'_{k+k'}|n_{k+k'+1},...,n_{k+k'+m},n'_{k+k'+m+1}...n'_{k+k'+m+m'}}\\
       =  \frac{(N - m-k)!}{N!}\frac{(N'-(m+k)N_q - k'-m')!}{(N'-(m+k)N_q)!}\sum^{'}\left(\prod_{\alpha = 1}^k e^{i n_\alpha \phi_\alpha} \prod_{\beta = k}^{k+k'} e^{i {n'_\beta} \psi_{\beta}}
       \prod_{\gamma = k+k'+1}^{k+k'+m} e^{-i n_\gamma \phi_\gamma} \prod_{\delta = k+k'+m+1}^{k+k'+m+m'} e^{-i {n'_\delta} \psi_{\delta}}\right)\\
   \end{aligned} 
\end{equation}
where we obtain the average by dividing by the total number of $m'+k'+m+k$-tuples with unique particles.
This result is corroborated by the computations done in Ref.~\cite{Voloshin:2010ut} for $\avg{1'+1}_{n'|n}$ and $\avg{1'+3}_{n',n|n,n}$.

\section{Evaluating $\avg{k'+m}$ using Q$_n$ Vectors }\label{ap:Qvec}

When defining $\avg{k'+m'+k+m}$ using a Q$_n$ vector method, we obtain a more complex recurrence relation, but also can weight each event differently in the calculation. More importantly, we are able to use far less computation time than it would take to make and iterate through $k'+m'+k+m$-tuples of particle angles. The event Q$_{n,t}$ vector is calculated by summing over all reference particle angles within the event:
\begin{equation}\label{Q}
    \text{Q}_{n,t} = \sum_{j=1}^N \omega_j^t e^{in\phi_j}
\end{equation}
with a weight, $\omega_j^t$, for each event. Currently $t$ is just a placeholder, but it will gain significance when considering autocorrelations within an event, and in most instances, it is unity. 
A $p_{n,t}$ vector, the POI analog to a Q$_n$ vector can be defined as: 
\begin{equation}
    \text{p}_{n,t} = \sum_{k = 1}^{N'} \omega_k^t e^{in\psi_k}
    \end{equation}
with the weighted sum over POI angles $\psi$. The overlap vector, $q_{n,t}$, can similarly be defined as:
\begin{equation}\label{q}
    \text{q}_{n,t} = \sum_{l = 1}^{N_q} \omega_l^t e^{in\theta_l}.
\end{equation}
with the weighted sum over the angles $\theta$ labelled as both POI and reference particles. Having defined the three different Q$_n$ vectors we will be using, we follow a recursive procedure of the same nature as outlined in \cite{Bilandzic:2013kga}. 
\begin{enumerate}
    \item We begin by writing the $\avg{k'+m'+k+m}$ correlator as a ratio of its numerator and denominator, where $D$ is a normalization factor and $N$ contains the actual correlations.  
    \begin{equation}
\avg{k'+m' + k + m} \equiv \frac{N\avg{k'+m' + k + m}}{D\avg{k'+m' + k + m}}
\end{equation}
    \item  We obtain the value of $\avg{k'+m'+k+m}$ first by considering an average over all $k'+m'+k+m$-tuples of POI and reference particle angles, in which each index references a unique particle. Additionally, we add a weight to each event, in keeping with Ref.~\cite{Bilandzic:2013kga}:
\begin{equation}
\begin{aligned}
   & N\avg{k'+m' + k + m}\equiv \\
   & \sum_{\substack{v_1, v_2, \ldots, v_{k'+m'+k+m} = 1 \\ v_1 \neq v_2 \neq \ldots \neq v_{k'+m'+k+m}}}^{N,N'} \left(\prod_{i=1}^{k} \omega_{v_i}e^{in_i\phi_{v_i}} \prod_{j=k}^{m+k} \omega_{v_j}e^{-in_j\phi_{v_j}}\prod_{h=m+k}^{k'+m+k} \omega_{v_h}e^{in_h\psi_{v_h}}\prod_{l = k'+m+k}^{k'+m'+k+m} \omega_{v_l}e^{-in_l\psi_{v_l}}\right) 
    \end{aligned}
\end{equation}
\begin{equation}
    D\avg{k'+m' + k + m} \equiv \sum_{\substack{v_1, v_2, \ldots, v_{k'+m'+k+m} = 1 \\ v_1 \neq v_2 \neq \ldots \neq v_{k'+m'+k+m}}}^{N,N'} \left(\prod_{i=1}^{k} \omega_{v_i} \prod_{j=k}^{m+k} \omega_{v_j}\prod_{h=m+k}^{k'+m+k} \omega_{v_h}\prod_{l = k'+m+k}^{k'+m'+k+m} \omega_{v_l}\right) 
\end{equation}
where the summation in both equations is over all groups of $k'+m'+k+m$ particles (of which $k'+m'$ are POI, and $k+m$ are reference particles), each with unique index $v_1,...,v_{k'+m'+k'+m}$. For each of the $k'+m'$ indices for POI particles, the index is allowed to run from one to $N'$, and for each of the $m+k$ indices for reference particles, the index is allowed to run from one to $N$.  
\item We can expand the previous equations, and substitute in the various event Q$_n$ vectors. 
We use the idea that the sum over each permutation of the product of weighted particle angles can be written as a product of the sums over each particle angle:
\begin{equation}\label{eq:taut}
  \sum_{r_1,...,r_{k'+m'+k+m}}^{N,N'}  \omega_{r_1}e^{in\phi_{r_1}} \cdots \omega_{r_{k'+m'+k+m}}e^{in\phi_{r_{k'+m'+k+m}}} = 
  \prod_{h = 1}^{k'+m'} \left(\sum_{i=1}^{N'} \omega_{i}e^{in_h\psi_{i}} \right)\prod_{j=1}^{k+m}\left(\sum_{l=1}^{N} \omega_{l}e^{in_j\phi_{l}}\right) 
\end{equation}
where the summation contains \textit{all} permutations of $r_1,...,r_{k'+m'+k+m}$, so some terms must be subtracted, accounting for the cases in which $r_j = r_k$ for two particle indices. Then, we substitute the Q$_n$ vectors introduced in Eq.~(\ref{Q}-\ref{q})
\begin{equation}
    \frac{N\avg{k'+m'+k+m}}{D\avg{k'+m'+k+m}}
     = \frac{\prod_{i=1}^{k+m} \text{Q}_{n_i,1} \prod_{j = 1}^{k'+m'} \text{p}_{n_j,1}}{\prod_{i=1}^{k+m} \text{Q}_{0,1} \prod_{j = 1}^{k'+m'} \text{p}_{0,1}} - \text{autocorrelations}  
\end{equation}
where the autocorrelations must be subtracted.
\item To account for autocorrelations, we consider the cases in which two indices reference the same particle angle. To do this, we recursively iterate through all $r$ "subcorrelators" with $k'+m'+k+m -1$ harmonics, from which $n_{k'+m'+k+m}$ is eliminated, and "added" onto the $r$-th harmonic:   
\begin{equation}
    \avg{k'+m'+k+m - 1}_{n_1, \cdots, (n_r + n_{k'+m'+k+m}) + \cdots + n_{k'+m'+k+m-1}}
\end{equation}
where this correlator considers the cases in which the angles from  particles in the $n_r$ and $n_{k'+m'+k+m}$ harmonic overlap. The recursive aspect to this algorithm means that we can apply the same procedure to calculate $\avg{k'+m'+k+m - 1}_{n_1, \cdots, (n_r + n_{k'+m'+k+m}) + \cdots + n_{k'+m'+k+m-1}}$, by calculating $\avg{k'+m'+k+m - 2}$, all the way until there are no more coinciding harmonics to consider, in the same manner as described in Ref.~\cite{Bilandzic:2013kga}.
\end{enumerate}

Using this recursive process, we can represent $\avg{k'+m'+k+m}$ using Q$_n$ vectors. However, in step 4.), it is not trivial to determine which Q$_n$ vector must be selected when representing the "overlapped" index $n_r = n_{k'+m'+k+m}$. In general, we consider four cases, regarding whether or not each harmonic is for POI or for reference particles. Note that in this paper, we've generally written the reference harmonics $n$ first, and the POI harmonics $n'$ second, but this ordering is arbitrary, and it is possible that any of these four cases might apply:
\begin{enumerate}[a.)] 
    \item $n_r$ and $n_{k'+m'+k+m}$ correspond to reference particle harmonics. In this case Q$_{n_r,1}$ is replaced by Q$_{n_r+n_{k'+m'+k+m},2}$ when evaluating $\avg{k'+m'+k+m- 1}_{n_1, \cdots, (n_r + n_{k'+m'+k+m}) + \cdots + n_{k'+m'+k+m-1}}$
    \item $n_r$ corresponds to a reference particle azimuthal angle, and $n_{k'+m'+k+m}$ corresponds to a POI angle. In this case, we replace  Q$_{n_r,1}$ with q$_{n_r+n_{k'+m'+k+m},2}$
    \item $n_r$ corresponds to a POI azimuthal angle, and $n_{k'+m'+k+m}$ corresponds to a reference particle angle. As before, we replace  p$_{n_r,1}$ with q$_{n_r+n_{k'+m'+k+m},2}$
    \item $n_r$ and $n_{k'+m'+k+m}$ both correspond to POI angles. In this case, we replace p$_{n_r,1}$ with p$_{n_r+n_{k'+m'+k+m},2}$.
\end{enumerate} 

We summarize the recurrence relation below, while noting that the evaluation of N$\avg{k'+m'+k+m-1}$ requires the use of the rules for each scenario detailed above:

\begin{eqnarray}
    N\avg{k'+m'+k+m}_{n_1, \cdots n_k, n'_{k+m+1}, \cdots , n'_{k+m+k'}|n_{k+1}\cdots n_{m},n_{k+m+k'+1}\cdots n_{k'+m'+k+m}}& &\nonumber\\
    =\prod_{i=1}^{k+m} \text{Q}_{n_i,1}\prod_{j=1}^{k'+m'} \text{p}_{n'_j,1} -  \sum_{r = 1}^{k'+m'+m+k-1} N \avg{k'+m'+k+m - 1}_{n_1, \cdots, (n_r + n_{k'+m'+k+m}) + \cdots + n_{k'+m'+k+m-1}}& &
\end{eqnarray}
Where we show that $N\avg{k'+m'+k+m}$ can be recursively evaluated by considering a  product of $k+m+k'+m$ for the Q$_n$ vectors, and then subtracting a sum of correlators with one fewer particle angle $N\avg{k'+m'+k+m-1}$, where for the $r$-th subtracted correlator, the $k'+m'+k+m$-th subscript $n_{k'+m'+k+m}$ is added to the $r$-th harmonic of the correlator.
A very simple example is shown below, where the two particle correlation between particles of interest and reference particles is written using Q$_n$ vectors: 
\begin{equation}
    \avg{1'+1}_{n,-n'} = \avg{e^{in(\phi_1 - \psi_2)}} = \frac{\text{Q}_{n,1}\text{p}_{n,1} - \text{q}_{2n,2}}{\text{Q}_{0,1}\text{p}_{0,1} - \text{q}_{0,2}}
\end{equation}
where this result is consistent with the same definition given in Ref.~\cite{Voloshin:2010ut}. There the weighting for each particle was uniform, and the denominator was rewritten in terms of $N,N'$ and $N_q$.

\section{Contribution of $v_n$, ${v'}_n$ to multi-POI cumulants}\label{ap:C}

We calculate the contribution of $v_n$ and $v'_n$ to the evaluation of $h_{n/p}^{k',m'}\{k'+m'+k+m\}$. We start with the definition of the generating function: 
\begin{equation}
    \mathcal{H}^{k',m'}_{n/p}(z) = \frac{\avg{e^{ip(\psi_1 + ... + \psi_{k'} - \psi_{k'+1} - ... - \psi_{k'+m'})}G_n(z)}}{\avg{G_n(z)}}
\end{equation}
that was previously discussed in Eq.\ (\ref{eq:H}).
Using the result from \cite{Borghini:2001vi,Bilandzic:2012wva}, we can determine the value of $\avg{G_n(z)}$ in terms of $v_n$, which approximates a Bessel function:
\begin{equation}
    \left\langle G_n(z)\right\rangle=\sum_{j=0,1,2, \ldots}^{M / 2} \frac{1}{M^{2 j}} \frac{M !}{(M-2 j) !(j !)^2} v_n^{2 j}|z|^{2 j} \approx I_0\left(2|z| v_n\right)
\end{equation}
We now calculate
$\avg{e^{ip(\psi_1 + ... + \psi_{k'} - \psi_{k'+1} - ... - \psi_{k'+m'})}G_n(z)}$ using the same method we used to evaluate $\mathcal{F}_{p/n}(z)$, by expressing the average as an integral around a fixed angle:
\begin{equation}
\avg{e^{ip(\psi_1 + ... + \psi_{k'} - \psi_{k'+1} - ... - \psi_{k'+m'})}G_n(z)}= \left.\frac{1}{2 \pi} \int_0^{2 \pi}\left\langle e^{ip(\psi_1 + ... + \psi_{k'} - \psi_{k'+1} - ... - \psi_{k'+m'})} G_n(z)\right\rangle\right|_{\Phi_n} d \Phi_n.
\end{equation}
Then, for every $e^{ip\psi_i}$, we substitute in $v'_p e^{ip\Phi_n}$, again following Ref.~\cite{Borghini:2001vi}, because the event averaged $\psi$ angles at $\Phi_n$ are not correlated with the event averaged $\phi$ angles at $\Phi_n$: 
\begin{equation}
\avg{e^{ip(\psi_1 + ... + \psi_{k'} - \psi_{k'+1} - ... - \psi_{k'+m'})}G_n(z)}= \frac{{v'_{p/n}}^{k'+m'}} {2\pi} \int_0^{2\pi} e^{ip(k'-m') \Phi_n} \langle G_n(z)\rangle |_{\Phi_n}d\Phi_n 
\end{equation}
We then solve the equation on the right, which only has nonzero solutions if $p(k'-m') = qn$ for some integer $q$. 
\begin{equation}
\int_0^{2\pi} e^{ip(k'-m') \Phi_n} \langle G_n(z)\rangle |_{\Phi_n}d\Phi_n 
 = \sum_{l=0}^{[(M+q) / 2]} \frac{M !}{(M-q-2 l) ! l !(2 l+q) !}\left(\frac{v_n}{M}\right)^{2 l+q} z^{* l} z^{l+q}
\end{equation}
Using the result from \cite{Borghini:2001vi}, we obtain the full expression for  $\avg{e^{ip(\psi_1 + ... + \psi_{k'} - \psi_{k'+1} - ... - \psi_{k'+m'})}G_n(z)}$ 
\begin{equation} 
\begin{aligned}
&\avg{e^{ip(\psi_1 + ... + \psi_{k'} - \psi_{k'+1} - ... - \psi_{k'+m'})}G_n(z)} \\
&= {v'_{p/n}}^{k'+m'} \sum_{l=0}^{[(M+\frac{p(k'+m)}{n}) / 2]} \frac{M !}{(M-\frac{p(k'+m)}{n}-2 l) ! l !(2 l+\frac{p(k'+m)}{n}) !}\left(\frac{v_n}{M}\right)^{2 l+\frac{p(k'+m)}{n}} z^{* l} z^{l+\frac{p(k'+m)}{n}}\\
&\approx {v'_{p/n}}^{k'+m} I_\frac{p(k'-m')}{n}\left(2|z| v_n\right)\left(\frac{z}{|z|}\right)^{\frac{p(k'-m')}{n}}
\end{aligned}
\end{equation}
Then, we obtain $\mathcal{H}^{k',m'}_{p/n}(z)$: 
\begin{equation}
    \mathcal{H}^{k',m'}_{p/n}(z) \approx \frac{I_\frac{p(k'-m')}{n}\left(2|z| v_n\right)}{I_0\left(2|z| v_n\right)}\left(\frac{z}{|z|}\right)^{\frac{p(k'-m')}{n}}{v'_{p/n}}^{k'+m'} 
\end{equation}
Note that this conclusion is consistent with $\mathcal{F}(z)$ as defined in Sec. \ref{gfc}, because letting $k' = m' = 1$ returns $\mathcal{F}_{n/n}(z) = {v'_n}^2$. Now that we have an expression for $\mathcal{H}^{k',m'}_{p/n}(z)$, we can also produce an expression $h_{p/n}\{k'+m'+k+m\}$, the power series expansion coefficients for $\mathcal{H}^{k',m'}_{p/n}(z)$, and express an estimate for ${v'_n}^{k'+m}$.

\section{Some properties of Central Moments}\label{ap:D}
Since central moments are not typically used to evaluate fluctuations in $v_n$, we demonstrate here some mathematical properties of the central moments that are similar to those of cumulants.
\begin{itemize}
    \item Central moments display \textbf{reduction}: if a stochastic variable $X_i$ is duplicated $j$ times, the resulting quantity is still a central moment, now of order $j$ in $X$: 
\begin{equation}
    \begin{aligned}
        \Tilde{\mu}_{\nu_1 ..., \nu_n}(\underbrace{X_1, \ldots, X_1}_j,..., X_n) = \avg{\prod_{i = 1}^j (X_1 - \avg{X_1})^{\nu_i} \prod_{i = j+1}^{n}(X_i - \avg{X_i})^{\nu_i}}\\
        = \avg{(X_1 - \avg{X_1})^{\sum_{i=1}^j \nu_i} \prod_{i = j+1}^{n}(X_i - \avg{X_i})^{\nu_i}}
        = \Tilde{\mu}_{(\nu_1 + ... + \nu_j)..., \nu_n}(X_1, ..., X_n)
    \end{aligned}
\end{equation}
\item Central moments display \textbf{translation invariance}: shifting the stochastic variable by some constant factor does not alter the value of $\Tilde{\mu}$. We use the fact that $\avg{X_i + c} = \avg{X_i} + c$:
\begin{equation}
    \begin{aligned}
     &\tilde{\mu}_{\nu_1,..,\nu_g..,\nu_n}(X_1..(X_g + c)..X_n) = \\
      &\avg {\left(\prod_{i = 1,i\neq g}^n (X_i - \avg{X_i})^{\nu_i} \right)(X_g +c - \avg{X_g + c})^{\nu_g}}  = \avg {\left(\prod_{i = 1,i\neq g}^n (X_i - \avg{X_i})^{\nu_i}\right) (X_i- \avg{X_g})^{\nu_g}} \\
      &= \tilde{\mu}_{\nu_1,..,\nu_g...,\nu_n}(X_1..X_g..X_n)
     \end{aligned}
\end{equation}
\item Central moments display \textbf{homogeneity}: they scale with their variables to their order. Using the fact that $\avg{cX}$ = $c\avg{X}$: 
\begin{equation}
    \begin{aligned}
        &\tilde{\mu}_{\nu_1,..,\nu_g..,\nu_n}(X_1..(cX_g)..X_n)\\
        &=\avg{\left(\prod_{i = 1,i\neq g}^n (X_i - \avg{X_i})^{\nu_i}\right)(cX_g - \avg{cX_g})^{\nu_g}} = \avg{\left(\prod_{i = 1,i\neq g}^n (X_i - \avg{X_i})^{\nu_i}\right)c^{\nu_g}(X_g - \avg{X_g})^{\nu_g}}\\
        &=c^{\nu_g}\tilde{\mu}_{\nu_1,..,\nu_g..,\nu_n}(X_1..X_g..X_n)
    \end{aligned}
\end{equation}
\end{itemize}
Additionally, central moments of order $\sum_i \nu_i \leq 3$ are identical to multivariate cumulants of the same order in each stochastic variable, meaning they obey some useful properties of cumulants, specifically additivity, and multilinearity: $\kappa_{n,...}(X + Y,...) = \kappa_{n,...}(X,...) + \kappa_{n,j}(Y,...)$.

\section{Explicit forms of Central Moments}
In Table~\ref{tab:2}, we introduce the first few central moments at orders $\nu = 2,3$ and $4$, for up to four unique stochastic variables $X,Y,Z$ and $W$, as defined in Sec.~\ref{vars}. Additionally, we briefly explain the various correlations and fluctuations measured by these central moments, and how they can be interpreted. The central moments with order $\nu = 2$ and $\nu = 3$ are also considered cumulants, as described in Sec.~\ref{sec:ASC}. 

\begin{sidewaystable}
    \caption{ We present the first few central moments, with different combinations of variables. }
    \begin{tabular}{c c c c c}
     \hline
     \hline
     \textbf{Central Moments} & \begin{tabular}{c}
          1 Stochastic Variable  \\
          (Univariate) 
     \end{tabular} & 2 Stochastic Variables & 3 Stochastic Variables & 4 stochastic variables \\ 
     
     $\nu = \sum_{i}\nu_i =  2$ &
     \begin{tabular}{c}
          $\avg{X^2} - \avg{X}^2$\\
          Variance: the fundamental\\
          measure of fluctuations.
     \end{tabular} 
      & \begin{tabular}{c}
          $\avg{XY}- \avg{X}\avg{Y}$\\
          Covariance: a measure of \\
          correlation between two variables,\\
          of which SC is a special case.\\
     \end{tabular} & & \\
     
     $\nu = 3$ & \begin{tabular}{c}
          $\avg{X^3} - 3\avg{X^2}\avg{X}$
          \\$ +2\avg{X}^3$\\
          Skewness: a  \\
         measure of a distribution's \\ 
         asymmetry.
          
     \end{tabular} 
     &
     \begin{tabular}{c} $\avg{X^2Y} - 2 \avg{XY}\avg{X}$\\ $-\avg{X^2}\avg{Y} +2\avg{X}^2\avg{Y}$ \\ Coskewness: \\ correlates linear deviations from \\
     the mean in $Y$ with nonlinear \\ (squared) deviations from the mean \\ in $X$.
     \end{tabular}
     &
     \begin{tabular}{c}
      $\avg{XYZ}-\avg{XY}\avg{Z}-\avg{XZ}\avg{Y}$ \\ $-\avg{YZ}\avg{X} +2\avg{X}\avg{Y}\avg{Z}$ \\
      Coskewness: a correlation \\ between deviations from the \\ mean in $X$ $Y$ and $Z$.
     \end{tabular}
    & \\
     
     $\nu = 4$ & \begin{tabular}{c}
          $\avg{X^4} - 4\avg{X^3}\avg{X}$ \\ $+6\avg{X^2}\avg{X}^2 - 3\avg{X}^4$\\
          Kurtosis: a \\
          measure of a distribution's \\
          tail "heaviness." 
     \end{tabular} 
     &
     \begin{tabular}{c}Cokurtosis in two variables can correlate \\ a deviation from the mean in\\ $Y$ to a deviation  from the \\ mean \textit{cubed} in $X$.\\
     $\avg{X^3Y} - 3\avg{X^2Y}\avg{X} $ \\ $-\avg{X^3}\avg{Y} + 3\avg{XY}\avg{X}^2 $ \\ $+3\avg{X^2}\avg{X}\avg{Y} - 3\avg{X}^3\avg{Y}$\\
     
     Cokurtosis can also correlate a  \\squared deviation
     from the mean in \\ $X$ to a squared
     deviation from \\the mean in $Y$.\\
    $\avg{X^2Y^2} - 2\avg{X^2Y}\avg{Y} -2\avg{XY^2}\avg{X} $ \\ $ +\avg{X^2}\avg{Y}^2 +4\avg{XY}\avg{X}\avg{Y} $ \\ $ +\avg{Y^2}\avg{X}^2 - 3\avg{X}^2\avg{Y}^2$\\
     \end{tabular}
      &
     \begin{tabular}{c}
     $\avg{X^2YZ}-\avg{X^2Y}\avg{Z}$ \\ $-\avg{X^2Z}\avg{Y}-2\avg{XYZ}\avg{X}$ \\ $ + 2\avg{XY}\avg{X}\avg{Z}+2\avg{XZ}\avg{X}\avg{Y} $ \\ $+ \avg{X^2}\avg{Y}\avg{Z} + \avg{YZ}\avg{X}^2$\\
     cokurtosis: correlating a squared \\ deviation from the mean in $X$ \\ to a deviation from the mean \\ in $Y$, and a deviation from \\ the mean in $Z$. 
     \end{tabular}
     &
     \begin{tabular}{c}
     $\avg{XYZW}-\avg{XYZ}\avg{W} -\avg{XYW}\avg{Z} $ \\ $-\avg{XZW}\avg{Y}-\avg{YZW}\avg{X}$ \\ $ + \avg{XY}\avg{Z}\avg{W}+\avg{XZ}\avg{Y}\avg{W} $ \\ $+\avg{XW}\avg{Y}\avg{Z} + \avg{YZ}\avg{X}\avg{W}$ \\ $+\avg{YW}\avg{X}\avg{Z}+\avg{ZW}\avg{X}\avg{Y}$ \\ $ - 3\avg{X}\avg{Y}\avg{Z}\avg{W}$ \\
     cokurtosis between 4 variables: \\
     the correlation between deviations \\ from the mean in $X$, $Y$, \\ $Z$, and $W$.
     \end{tabular}\\
     \hline
     \hline
     \end{tabular}
     \label{tab:2}

\end{sidewaystable}

\end{document}